\documentclass[sn-basic]{sn-jnl}

\usepackage{graphicx}
\usepackage{xspace}
\usepackage{xcolor,array}
\usepackage{booktabs}
\usepackage{amsmath}
\usepackage{url}
\usepackage{doi}
\usepackage{pgfgantt}
\usetikzlibrary{shapes.misc}

\newcommand{\tOne}[1]{\textcolor{blue!70!black}{#1}}
\newcommand{\tTwo}[1]{\textcolor{red!70!black}{#1}}
\newcommand{\scell}[2]{
  \raisebox{-1ex}[1ex]{%
  \shortstack[c]{%
    \rule{0pt}{2ex}\tOne{#1}\\
    \rule{0pt}{1.5ex}\tTwo{#2}
  }%
}
}
\usepackage[table]{xcolor} % Add this in the preamble
\definecolor{lightblue}{HTML}{D9EAF7} % Light blue shade
\definecolor{foobarblue}{RGB}{0,153,255}
\usepackage{colortbl}

\newcommand{\chatgpt}{\textsc{ChatGPT}\xspace}
\newcommand{\jme}{\textsc{jMonkeyEngine}\xspace}
\newcommand{\battleship}{\textit{Battleship}\xspace}
\newcommand{\preparation}{\textit{Preparation}\xspace}
\newcommand{\design}{\textit{Design}\xspace}
\newcommand{\analysis}{\textit{Analysis}\xspace}
\newcommand{\coding}{\textit{Coding \& Testing}\xspace}

\begin{document}

\title[Learning to Code with Context]{Learning to Code with Context:  A Study-Based Approach}

\author*[]{\fnm{Uwe M.} \sur{Borghoff}}\email{uwe.borghoff@unibw.de}
%\orcid{0000-0002-7688-2367}
\author*[]{\fnm{Mark}    \sur{Minas}}\email{mark.minas@unibw.de}
%\orcid{0000-0002-8968-9013}
\author*[]{\fnm{Jannis}  \sur{Schopp}}\email{jannis.schopp@unibw.de}
%\orcid{0009-0006-0776-9156}

\affil[]{\orgdiv{Institute for Software Technology, Department of Computer Science}, \orgname{University of the Bundeswehr Munich}, \orgaddress{\street{Werner-Heisenberg-Weg 39}, \city{85579 Neubiberg}, \country{Germany}}}

%%%%%%%%%%%%%%%%%%%%%%%%%%%%%%%%%%%%%%%%%%%%%%%%%%%%%
\abstract{The rapid emergence of generative AI tools is transforming the way software is developed. Consequently, software engineering education must adapt to ensure that students not only learn traditional development methods but also understand how to meaningfully and responsibly use these new technologies. In particular, project-based courses offer an effective environment to explore and evaluate the integration of AI assistance into real-world development practices.
This paper presents our approach and a user study conducted within a university programming project in which students collaboratively developed computer games. The study investigates how participants used generative AI tools throughout different phases of the software development process, identifies the types of tasks where such tools were most effective, and analyzes the challenges students encountered. 
Building on these insights, we further examine a repository-aware, locally deployed large language model (LLM) assistant designed to provide project-contextualized support. The system employs Retrieval-Augmented Generation (RAG) to ground responses in relevant documentation and source code, enabling qualitative analysis of model behavior, parameter sensitivity, and common failure modes. The findings deepen our understanding of context-aware AI support in educational software projects and inform future integration of AI-based assistance into software engineering curricula.
}


\maketitle

%%%%%%%%%%%%%%%%%%%%%%%%%%%%%%%%%%%%%%%%%%%%%%%%%%%%%

%%%%%%%%%%%%%%%%%%%%%%%%%%%%%%%%%%%%%%%%%%%%%%%%%%%%%
\section{Introduction}\label{s:intro}
%%%%%%%%%%%%%%%%%%%%%%%%%%%%%%%%%%%%%%%%%%%%%%%%%%%%%
The programming project at our university is a software development course in which group work in a software development team is a crucial aspect. Similar courses are probably part of almost all bachelor programs in computer science worldwide.
In our programming project, about seven students work together in a team to create a computer game using a structured development process. The main goal is for the participants to gain experience in software development in a team with all the necessary activities, including planning, programming, testing and documentation, as well as the necessary soft skills such as teamwork and communication. 

The general availability of generative AI tools, such as \chatgpt and GitHub’s \textsc{Copilot} \citep{MastropaoloPGCSOB23}, also has an impact on the way software is developed. So far, such AI tools are not part of the curriculum of our programming project. When asked personally, our students naturally say that they use them frequently, but are often dissatisfied with their answers. To counteract this uncontrolled use of AI and to train our students in its responsible use, we want to provide participants in our programming project with an AI tutor in the coming years. The AI tutor will be chat-based and will support students in software development and, in particular, in the creation of programs. To counteract the much-discussed effect of \textit{skill degradation} \citep{CrowstonB25,macnamara2024does}, however, it will not simply present ready-made solutions, but will provide hints for solutions like a human tutor and guide students to solve problems independently. 

In recent months, we have followed an empirical approach to plan the implementation and use of the AI tutor. This paper\footnote{This paper builds on and extends the work presented by \cite{BorghoffMS25}, which is distributed under the Creative Commons Attribution International~4.0~License. Compared to our conference paper (Sect.~\ref{s:project}--\ref{s:userstudyresults}), this journal version introduces new Sect.~\ref{s:ext-design}--\ref{s:ext-results} that present a repository-aware, locally deployed LLM assistant, as well as a qualitative analysis of its behavior in a programming project setting. This extension is a direct follow-up on the results of our user study, addressing the identified limitations of general-purpose AI tools for code comprehension and integration within software development projects.} describes our multi-stage approach and the results we have achieved with it.

In a \textit{first step}, we empirically collected data on how our students use freely available AI tools. To this end, we conducted a user study among the participants in last year's programming project from October to December 2024. From the outset, they were asked to understand freely available AI tools as useful aids and to use them as such, e.g., for coding or documenting. At the end of the programming project, we conducted a study in the form of a structured questionnaire that asked about their experiences with AI tools in the various phases and in relation to the various activities. 

As a result of this user study, we were able to identify several shortcomings of the freely available AI tools in this application domain. Among other things, they are not familiar with the documents and source code of the sample program that are made available to the participants and that they need to familiarize themselves with. The AI tools were therefore almost never able to provide helpful answers when asked questions by students in this context. To counteract this, our own AI tutor must therefore have access to relevant documents and the source code of the project and use them appropriately.

In the \textit{second step} of our empirical approach, we conducted a case study in which we implemented a repository-aware, locally deployed large language model (LLM) assistant that offers project-contextualized support. The system employs RAG to ground responses in relevant documentation and source code. With this assistant, we were able to investigate model behavior, parameter sensitivity, and common failure modes. In order to evaluate the correctness of the answers produced, the assistant in this case study---in contrast to our planned AI tutor---still responded with ready-made solutions. 

An important aspect of these investigations was that we also have to operate our planned AI tutor locally for data protection reasons. In order to guarantee response times from the AI tutors even when used by several students at the same time, the LLM used should therefore be as small as possible. The case study was therefore intended to answer the question of which LLMs can correctly answer the selected programming queries and how well smaller LLMs with fewer parameters are able to do so. 

The rest of the paper is organized as follows. After reviewing related work in Sect.~\ref{s:related-work}, we outline the context of the study, focusing on our programming project in Sect.~\ref{s:project}. Sections~\ref{s:study}--\ref{s:userstudyresults} describe our user study and its results, while Sections~\ref{s:ext-design}--\ref{s:ext-results} present our case study and its results. Specifically, Section~\ref{s:study} describes the demographics and methodology of the user study, and Section~\ref{s:userstudyresults} summarizes the quantitative findings. Next, we introduce our case study: a repository-aware, locally deployed assistant. Section~\ref{s:ext-design} details the system design and local deployment. Section~\ref{s:ext-Study} explains the qualitative methods and data capture. Section~\ref{s:ext-results}  reports the qualitative results, including parameter sensitivity, model comparison, and failure modes. We conclude in Sect.~\ref{s:conclusion}.

%%%%%%%%%%%%%%%%%%%%%%%%%%%%%%%%%%%%%%%%%%%%%%%%%%%%%
\section{Related Work}\label{s:related-work}
%%%%%%%%%%%%%%%%%%%%%%%%%%%%%%%%%%%%%%%%%%%%%%%%%%%%%
\cite{Rane2023} examine the use of AI technologies, such as machine learning and natural language processing, to create adaptive learning environments that adjust instructional strategies based on learner feedback and progress. 
\cite{NitzlCKB25} similarly show how AI can enhance performance in the intelligence analysis phase.

The use of AI, particularly conversational and generative models, has been widely studied in software engineering \citep{Russo2024,SengulND24}. 
A comprehensive overview is provided by \cite{Kokol24}, and a dedicated special issue on generative AI in software engineering is presented by \cite{CarletonFZX24}. 
Recent studies confirm that AI and deep learning are becoming key components of modern software engineering practice \citep{Yang2022}, showing measurable gains in efficiency, accuracy, and productivity \citep{MartinovicR25,Nascimento2023,PiscitelliCRF24,WaseemDA0FM24}. 
\cite{Bhandari2023} further highlight emerging intersections between AI and software engineering that promise to reshape both research and education.

Collectively, these works indicate that integrating AI into software projects has significant implications for learning and professional skill development. While AI can automate routine tasks and increase productivity, it also calls for new competencies and balanced human–AI collaboration strategies.

The position paper by \cite{Daun2023} discusses \chatgpt's potential to generate natural language explanations and provide personalized tutoring for software engineering students. Likewise, \cite{BorghoffMM24} demonstrate how automated tools can support program evaluation, grading, and code generation, streamlining assessment processes and improving feedback quality.

Automatic assessment systems, or \textit{autograders} \citep{AlaMutka05}, have long supported programming education. Test-based systems execute student submissions against instructor-defined cases, a practice dating back to the 1960s \citep{DouceLO05}. Contemporary examples include \textsc{ArTEMiS}\footnote{\url{https://github.com/ls1intum/ArTEMiS}} \citep{KruscheS18}, which offers an online coding environment with immediate feedback, and \textsc{Praktomat}\footnote{\url{https://www.waxmann.com/automatisiertebewertung/}}, which supports Java and \textsc{JUnit}-based testing while integrating plagiarism detection via \textsc{JPlag}\footnote{\url{https://github.com/jplag/JPlag}} and style checking through \textsc{Checkstyle}\footnote{\url{https://checkstyle.org/}}. Newer systems employ machine learning for automatic grading \citep{Srikant:13} and adaptive evaluation \citep{ChrysafiadiVT22,Hidalgo-SuarezB22}.

With the advent of deep learning and \textit{Large Language Models} (LLMs) capable of generating code from extensive training corpora, new opportunities have emerged. Models such as DeepMind’s \textsc{AlphaCode} \citep{Li2022}, Google’s PaLM \citep{ChowdheryNDBMRBCSGSSTMRBTSPRDHPBAI23}, Microsoft’s \textsc{CodeBERT} \citep{FengGTDFGS0LJZ20}, and OpenAI’s \textsc{ChatGPT} and \textsc{Codex} \citep{Finnie-AnsleyDB22} can produce syntactically correct and functionally meaningful code from textual prompts. These advances have inspired a growing ecosystem of AI-based development assistants that improve code quality, security, and accessibility for novice programmers \citep{KazemitabaarCME23}. 

Table~\ref{table:aisystems_extended} summarizes representative systems used in educational and professional settings.

\begin{table}[ht]
\centering
%\begin{threeparttable}
\caption{Representative AI-based systems for code generation and assistance.}
\label{table:aisystems_extended}
\begin{tabular}{p{3.1cm}p{7.8cm}}
\hline
\rowcolor{lightblue}
\raisebox{-2mm}[2mm]{\textbf{System}} & \raisebox{-2mm}[2mm]{\textbf{Key Features}} \\[4mm]
\hline\\[-2mm]
\textsc{AlphaCode}\tnote{a} & DeepMind’s competitive-programming model generating solutions from natural-language problem statements. \\
\textsc{PaLM}\tnote{b} & Google’s large language model capable of reasoning and text-to-code generation across domains. \\
\textsc{CodeBERT}\tnote{c} & Transformer pretrained on paired natural language and source code for code understanding and synthesis. \\
\textsc{ChatGPT}/\textsc{Codex}\tnote{d} & OpenAI’s conversational and code-oriented LLMs for generating, explaining, and refactoring code. \\
\textsc{CodeWhisperer}\tnote{e} & Amazon’s AI assistant providing full-function suggestions and built-in security checks in real time. \\
\textsc{GhostWriter}\tnote{f} & Replit’s in-editor assistant offering code completion, transformation and inline explanations. \\
\textsc{Codota/Tabnine}\tnote{g} & Predictive code-completion tools trained on open-source corpora. \\
\textsc{Cody}\tnote{h} & Sourcegraph’s context-aware assistant supporting repository navigation and refactoring. \\
\textsc{Diffblue Cover}\tnote{i} & AI-driven test-generation tool for Java code, automating unit test creation and improving coverage. \\
\textsc{Qodo}\tnote{j} \, (aka \textsc{CodiumAI}) & AI platform focused on code generation and quality assurance, with context-aware review and test scaffolding features. \\
\textsc{Cursor AI}\tnote{k} & Coding assistant indexing code bases and enabling natural-language prompts for code generation and modification. \\
\hline
\end{tabular}
\begin{tablenotes}
\footnotesize
\item[a] \url{https://alphacode.deepmind.com/} (all accessed November 20, 2025)
\item[b] \url{https://ai.google/discover/palm2/}
\item[c] \url{https://github.com/microsoft/CodeBERT}
\item[d] \url{https://openai.com/de-DE/index/introducing-codex/}
\item[e] \url{https://workshops.aws/categories/CodeWhisperer}
\item[f] \url{https://blog.replit.com/ai}
\item[g] \url{https://www.tabnine.com/blog/codota-is-now-tabnine/}
\item[h] \url{https://about.sourcegraph.com/cody}
\item[i] \url{https://www.diffblue.com/}
\item[j] \url{https://qodo.ai/}
\item[k] \url{https://cursor.com/}
\end{tablenotes}
%\end{threeparttable}
\end{table}

\cite{gao2024} compare retrieval-augmented generation (RAG) with fine-tuning and prompt engineering, demonstrating that RAG integrates external knowledge more effectively and adapts more readily to new contexts.
Similarly, \cite{OvadiaBME24} find that RAG outperforms unsupervised fine-tuning by combining existing information with newly retrieved data.
Together, these studies emphasize the central role of contextual awareness, which underpins the repository-based approach adopted in this work.
Additional perspectives on this topic are provided in the recent surveys by  \cite{Huang2024} and \cite{NiEtAl2025}.

These developments also create new opportunities to leverage large language models for improving code quality and supporting novice programmers \citep{KazemitabaarCME23}.
For example, experiments by \cite{ManakinaL25} demonstrate that LLMs can serve as effective tools for software engineering curriculum design, maintaining instructional quality and offering structured assignment guidance to students.
Many programming tasks that once required human expertise can now be performed by AI systems, either autonomously or in collaboration with humans.
\cite{RamanK22} analyze the coding process through a six-step framework, evaluating automation tools such as GitHub’s \textsc{Copilot} \citep{MastropaoloPGCSOB23}, and conclude that future software engineering education should emphasize code comprehension over coding itself \citep{IzuSACDGHKLMW19a}.
\cite{Moroz2022} further observe that, when using \textsc{Copilot}, the goal is often not to obtain a single correct solution but to explore multiple viable strategies.

However, other studies have also reported drawbacks and negative experiences.
\cite{Choudhuri2024} found no significant improvement in productivity or self-efficacy when using \chatgpt compared with traditional resources, yet participants reported greater frustration, citing limited advice on specialized topics, incomplete assistance, and fabricated responses.
Similarly, \cite{Pudari2023} note that although syntax-level support has reached a mature stage, abstract aspects such as idiomatic language use and complex design principles remain beyond the current capabilities of AI tools.
Quantitative and qualitative analyses by \cite{AlOmar25} reveal that students remain skeptical about using \chatgpt for code reviews while also suggesting ways to improve review practices and raise awareness of LLMs in student software development projects.

\cite{li2025generative} show that, while generative AI can be helpful for programming education, excessive dependence on it can hinder long-term knowledge transfer.
\cite{Garousi2025} also identify concerns about overreliance, reduced critical thinking, and long-term skill deficits in software engineering education.

%%%%%%%%%%%%%%%%%%%%%%%%%%%%%%%%%%%%%%%%%%%%%%%%%%%%%
\section{The Programming Project}\label{s:project} 
%%%%%%%%%%%%%%%%%%%%%%%%%%%%%%%%%%%%%%%%%%%%%%%%%%%%%
For most of our students, the programming project is their first software development project, where they work in a team of about seven students and realize a complete computer game in Java. The syllabus for this course is worth 9 credits, which corresponds to a workload of about 270 hours in the period from October to December.

We have offered and improved this course for many years. The following paragraphs describe the organisation of this and the previous year (i.e., 2024 and 2025).

%%%%%%%%%%%%%%%%%%%%%%%%%%%%%%%%%%%%%%%%%%%%%%%%%%%%%
\subsection{Basic Organization}
%%%%%%%%%%%%%%%%%%%%%%%%%%%%%%%%%%%%%%%%%%%%%%%%%%%%%
Each team is assigned a tutor and a supervisor. The tutors are students who took the course the previous year, and they assist the team members with the usual problems encountered in a software development project. The supervisors are members of the academic staff who provide higher-level support to the teams, monitor and evaluate the teams' progress, approve artifacts, and control when the teams are allowed to proceed to the next phase and when the project is successfully completed.

Prior to the course, a list of twelve game proposals is compiled, each of which is sketched on a page by the tutors. All games are multiplayer 3D games with network support, based on popular board and card games (e.g., Monopoly, Ludo, Risk), and about equally difficult to implement. Students enrolled in the course are asked to form seven teams and prioritize a list of three games from this list of all games before the course begins. Based on these priorities, games are assigned to these teams so that each team realizes a different game.

The course then begins with the \textit{kick-off meeting}, where all students, supervisors, and tutors meet in the lecture hall to discuss organizational issues. In the following weeks, each team works independently on their projects, following the traditional waterfall process model as described below. 
This process model was deliberately chosen for three reasons. First, it is simple and clearly structured, and it is the first software development project in a team for most of our students. As a result, they are unfamiliar with the different phases and artifacts involved in software development. 
Second, the course lasts only 11 weeks, which is very little time to learn and try out a more modern, agile process model.
Finally, and most importantly, the more rigid waterfall setup makes it easier to measure the students' use of AI in concrete parts of the project.

At the end of the term, the teams present their games to the other teams and invited guests, e.g., other students, in a so-called \textit{public game presentation}. This event also includes a competition where each participant (including all guests) has the opportunity to rate each game presented, and the winning team is awarded a prize.

\begin{figure}[tb]
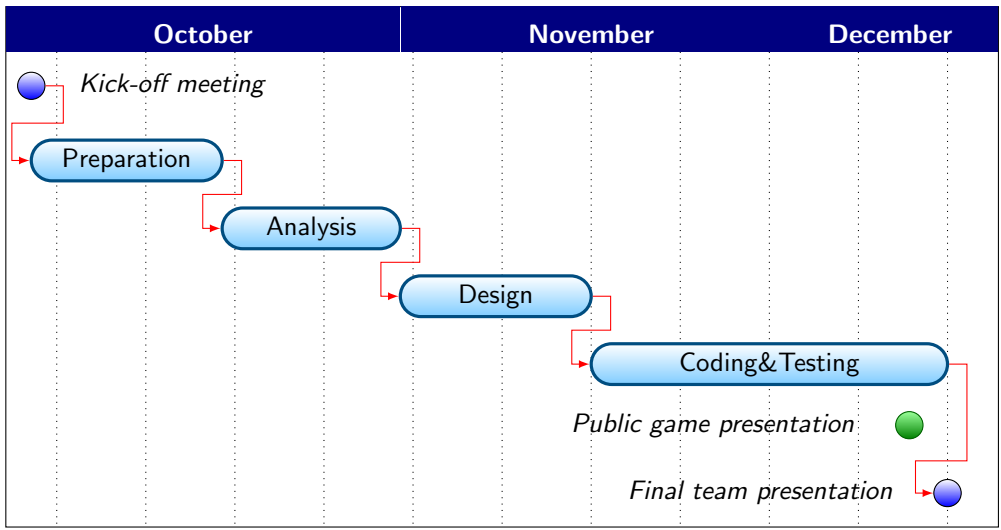

\begin{ganttchart}[
expand chart=\textwidth,
%x unit=0.12cm,
y unit title=.6cm,
y unit chart=.9cm,
vgrid={*3{draw=none}, dotted, *3{draw=none}},
time slot format=isodate,
link bulge=1.4,
inline,
link/.style={-latex, draw=red},
title/.append style={draw=none, fill=blue!50!black},
title label font=\sffamily\bfseries\color{white},
title label node/.append style={below=-1.6ex},
title left shift=.05,
title right shift=-.05,
title height=1,
bar/.append style={
shape=rounded rectangle,
inner sep=0pt,
draw=foobarblue!50!black,
very thick,
top color=white,
bottom color=foobarblue!50
},
bar height=.6,
bar label font=\sffamily\color{black},
milestone label font=\sffamily\slshape\color{black},
milestone/.append style={shape=circle,inner sep=3pt,top color=white,bottom color=blue,draw=black},
milestone inline label node/.append style={left=6mm},
]{2024-10-01}{2024-12-17}
\gantttitlecalendar{month=name} \\
\ganttmilestone[milestone inline label node/.append style={right=5mm}]{Kick-off meeting}{2024-10-02} \\
\ganttbar{Preparation}{2024-10-03}{2024-10-17} \\
\ganttbar{Analysis}{2024-10-18}{2024-10-31} \\
\ganttbar{Design}{2024-11-01}{2024-11-15} \\
\ganttbar{Coding\&Testing}{2024-11-16}{2024-12-13} \\
\ganttmilestone[milestone/.append style={draw=green!40!black, top color=green!40!white, bottom color=green!50!black}]{Public game presentation}{2024-12-10} \\
\ganttmilestone{Final team presentation}{2024-12-13}
\ganttlink{elem0}{elem1}
\ganttlink{elem1}{elem2}
\ganttlink{elem2}{elem3}
\ganttlink{elem3}{elem4}
\ganttlink[link mid=.75]{elem4}{elem6}
\end{ganttchart}
  \caption{Schedule of the programming project.}
  \label{f:project}
\end{figure}

The project schedule for each team is organized into four sequential phases as shown in Fig.~\ref{f:project} and explained below. 

Each team has a fixed weekly one-hour meeting with the supervisor and tutor, where team members present their progress and discuss problems and solutions with the supervisor and tutor.  Each phase ends at a milestone, and teams are not allowed to proceed to the next phase until they have successfully completed the previous phase. It is the supervisor's responsibility to evaluate and approve the team's progress and artifacts. As a result, each team's individual phases can take longer than planned, causing teams to fall behind schedule. Last year, this was the case for some teams  (especially in the \design phase), but all were able to catch up by the end of the term and participate in the \textit{public game presentation} with an executable game.

%%%%%%%%%%%%%%%%%%%%%%%%%%%%%%%%%%%%%%%%%%%%%%%%%%%%%
\subsection{Project Phases}
%%%%%%%%%%%%%%%%%%%%%%%%%%%%%%%%%%%%%%%%%%%%%%%%%%%%%
The phases are as follows.
The first phase is the \preparation phase (2 weeks), during which each team member works on training tasks and practice programming techniques. This phase is independent of the specific games that the teams would implement in the later phases. Instead, a realization of the simple \battleship game is available to all teams from the start. Like the games to be realized later, it is a multiplayer 3d game, playable over the network, and implemented on top of the freely available 3d game engine \jme. \battleship was designed to demonstrate the architecture and various concepts of 3d games, and also serves as a starting point and blueprint for the teams' own game development. The task of the \preparation phase is therefore to familiarize the students with \battleship and its structure and to extend it.

The following three phases follow the waterfall process model. During the \analysis phase (2 weeks), the teams create a product requirement specification based on the one-page sketch of their game. The teams produce artifacts such as an extended game description as prose text, a data model described by a class diagram, a list of all use cases and their descriptions based on a use case template, a description of acceptance tests as a table of test cases, and a first version of the user manual including sketches of the GUI.

The software specification document is created by the teams in the following \design phase (2 weeks) and consists of a prose description of the game specification, in particular the chosen architecture and the communication protocol of the required network protocol. Recommended diagrams for specifying details are, e.g., class diagrams, BPMN diagrams, sequence diagrams and state diagrams. System, integration and unit tests are specified in a test case table.
In the \coding phase (about 3 weeks), the games are finally implemented and tested according to the teams' software specification documents. Any deviations from these documents are required to be documented, and the final version of the user manual is made available.

A team is considered to have successfully completed their programming course when their supervisor has approved their final product and they have presented their project and game to the supervisor in a \textit{final team presentation}. Upon meeting these criteria, all team members are awarded the course credits.

%%%%%%%%%%%%%%%%%%%%%%%%%%%%%%%%%%%%%%%%%%%%%%%%%%%%%
\subsection{Course Evolution}
%%%%%%%%%%%%%%%%%%%%%%%%%%%%%%%%%%%%%%%%%%%%%%%%%%%%%
The course of our programming project has evolved over the years, but we have kept its basic organization with teams realizing games, using Java and the waterfall process model, and presenting all games in the \textit{public game presentation}. The latter has proven to be an essential aspect of the programming project for the following reason: Participants do not receive a grade at the end of the course because it is difficult to evaluate the individual performance of the participants. Therefore, only successful participation is confirmed. Without the ``reward'' of a better grade, there is less motivation to commit to the project beyond the minimum effort. The public presentation of one's own game and, above all, the competition between the teams for the prize of the best, most beautiful and funniest game counteracts this. This is evident from the fact that almost all teams have finished their game in time for the public presentation date in every run of the programming project, even when they were behind schedule due to one of the phases taking longer than planned. In addition, many students stated that the public presentation of their game was a highlight of the course and was fun.

All of the major evolutionary steps over the years have been in response to external events. 

The first major evolutionary step was triggered by the Covid pandemic. The entire course and teamwork had to be done online. We had to respond to user feedback as described by \cite{BorghoffMM23}. Unlike before, all teams had to make the same game (a dungeon crawler in 2022 and a racing game in 2023), but with different themes. To prevent the teams from sharing their solutions, all participants had to complete assignments that were evaluated by an automatic program assessment system. 

With the end of the pandemic, we abandoned the assignments and had different games realized. This decision was also influenced by the fact that the teams spent a lot of time solving the assignments, which meant that the actual project work and the quality of the games suffered. 
 
In the evaluation of the last term in 2024, a large number of students confirmed that they preferred to return to the programming project without assignments and that it was more fun that way. In fact, the quality of the 2024 games has increased significantly compared to the two years with assignments, considering the number of implemented functions and the graphical design of the games.

There is also a growing trend among students to adopt generative AI tools in their workflows. Although these tools are not formally integrated into the course curriculum, they have been encouraged for specific tasks, such as code documentation and text proofreading, in both last year's and this year's course.

%%%%%%%%%%%%%%%%%%%%%%%%%%%%%%%%%%%%%%%%%%%%%%%%%%%%%
\section{User Study}\label{s:study}
%%%%%%%%%%%%%%%%%%%%%%%%%%%%%%%%%%%%%%%%%%%%%%%%%%%%%
At the end of the 2024 term, a user study was conducted during the final presentations to investigate the extent and nature of AI tool usage. Understanding how students adopt and apply general-purpose AI tools, mostly large language models, during the programming project can provide valuable insights for developing dedicated AI tools to better support our software engineering course. However, these tools are neither formally integrated into the course curriculum nor openly discussed by students, highlighting the need for further investigation to assess their usage and impact.

Rather than advocating a specific approach to AI integration we explore how students naturally incorporate AI tools into a software development project. We aim to establish a baseline understanding of AI adoption in order to determine what types of support, tutoring or other AI-driven interventions may be needed in the future.

A recent experiment by \cite{RapakaDKMCG25} examines two versions of an educational game. One version followed a traditional design, while the other incorporated an intelligent, AI-based process. The experiment demonstrates that immersive and AI technologies can serve as valuable tools in the development of educational games and entertainment applications.
At the same time, previous research by \cite{Choudhuri2024} highlights that using AI in learning software engineering can significantly increase frustration levels.

To address these questions, we conducted a user study during the October-December term of 2024 to capture a snapshot of AI adoption among the student body. The goal of the study was to determine which aspects of the course students engaged with using AI tools, identify successful applications, and uncover any remaining challenges.

%%%%%%%%%%%%%%%%%%%%%%%%%%%%%%%%%%%%%%%%%%%%%%%%%%%%%
\subsection{Demographics}
%%%%%%%%%%%%%%%%%%%%%%%%%%%%%%%%%%%%%%%%%%%%%%%%%%%%%
The study included 38 participants, representing approximately 78\% of the 49 students enrolled in the course. As shown in Table~\ref{tab:demographics}, most of the participants were Computer Science students, followed by Business Informatics and Mathematical Engineering students. The gender and academic program distribution of the study participants closely mirrored that of the overall course population.

\begin{table*}[t]
\caption{Demographics of course and study participants.}
\centering
\setlength{\tabcolsep}{6pt}
\renewcommand{\arraystretch}{1.30}
\begin{tabular}{lcc}
\hline
\rowcolor{lightblue}
 & \textbf{Course participants} & \textbf{Study participants} \\ 
 \rowcolor{lightblue} 
\textbf{Characteristic} & \textbf{(n=49)} & \textbf{(n=38)}   \\ \hline
\multicolumn{3}{l}{\textit{Gender}}                           \\
\hspace{5mm}Male                        &   46  &   35        \\
\hspace{5mm}Female                      &   3   &   3         \\[2mm] 
\multicolumn{3}{l}{\textit{Academic programs}}                \\
\hspace{5mm}Computer Science            &   34  &   26        \\
\hspace{5mm}Business Informatics        &   13  &   10        \\ 
\hspace{5mm}Mathematical Engineering    &   2   &   2         \\[2mm] 
\textit{Participation rate} &   --   &   $\approx$ 78\%       \\\hline
\end{tabular}
\label{tab:demographics}
\end{table*}

Of the 49 students, 43 were in the middle of their bachelor's program, while 6 were nearing its completion. At the time of the study, all participants had completed the required undergraduate courses in computer science and object-oriented programming.

%%%%%%%%%%%%%%%%%%%%%%%%%%%%%%%%%%%%%%%%%%%%%%%%%%%%%
\subsection{Methodology}
%%%%%%%%%%%%%%%%%%%%%%%%%%%%%%%%%%%%%%%%%%%%%%%%%%%%%
Data for the study were collected using a structured questionnaire distributed during the final team presentation of the programming project. This timing allowed participants to reflect on their experiences during all phases of the project. 

Participation in the study was voluntary, and students were informed of their right to withdraw at any time. Anonymity of data was ensured, and participants were given the opportunity to request that their data be deleted. The study adhered to general data protection regulations to ensure the ethical handling and protection of personal data.

The questionnaire was structured to comprehensively evaluate the adoption and application of AI tools during the programming project, following the \textit{Technology Acceptance Model} \citep{davis1989}. This framework emphasizes factors that influence technology acceptance, such as perceived usefulness and ease of use.

%%%%%%%%%%%%%%%%%%%%%%%%%%%%%%%%%%%%%%%%%%%%%%%%%%%%%
\subsection{Questionnaire}
%%%%%%%%%%%%%%%%%%%%%%%%%%%%%%%%%%%%%%%%%%%%%%%%%%%%%
We invited all attending students to participate in the study, regardless of whether or not they used AI tools. Including students who did not use AI was essential, as their insights provide valuable context for building an authentic picture of AI adoption. For those who did not use AI, we asked them to share their reasons for not using AI, providing perspective on barriers to adoption or other influencing factors.

\begin{enumerate}

    \item \textit{General question:}
        \newline
        \textit{``Did you use AI tools during your programming project?''}
            
            Options: Yes, No. If ``No'', participants provided reasons such as lack of knowledge about available tools, no access to suitable tools, privacy or trust concerns, and other (open-ended response).
               
\end{enumerate}

Perceived usefulness was incorporated into the study by asking students to rate how AI tools supported their tasks throughout the programming project phases, which included \preparation, \analysis, \design, and \coding. 
The questionnaire captured several dimensions that are helpful in understanding the adoption and use of generative AI tools. Participants were asked to describe the tasks they completed using AI tools and to identify the platforms they used.
Questions focused on the effectiveness of AI in
Debugging, Documentation, and Code Generation, gauging how well these tools contributed to student success and efficiency in specific project contexts.

\begin{enumerate}
    \setcounter{enumi}{1} 

    \item \textit{Repeated questions for each project phase:}
        \begin{enumerate}
            \item \textit{``Did you use AI tools during this phase?''}
            
            Options: Yes, No.
            \item \textit{``What specific AI platforms did you use during this phase?''} 
            
            Examples: \chatgpt, \textsc{Copilot}, \textsc{Claude}, \textsc{Gemini}, \textsc{LLaMA}, or other (open-ended response).
            \item \textit{``What tasks have you worked on using AI tools?''}
            
            Participants could choose from predefined tasks or provide their own description in an open-ended response.
            \item \textit{``How helpful was AI in this phase?''}  
            
            Scale: 1 (not helpful) to 5 (very helpful).
            \item \textit{``How much time did the AI save you during this phase?''}
            
            Options: None, Less than 1 hour, 1–-3 hours, 4–-6 hours, 7–-9 hours, 10+ hours.
            \item \textit{``Did you face challenges when using AI tools?''}
            
            Options: Yes, No. If ``Yes'', participants could specify challenges such as unclear results, difficulty integrating tools, and other (open-ended response).
        \end{enumerate}

\end{enumerate}

Responses on the helpfulness were recorded using a five-point Likert scale ranging from ``not helpful'' to ``very helpful''. Ease of use was addressed by exploring the challenges of integrating AI tools into the workflow. Open-ended questions invited participants to describe barriers to usability, such as unclear results. This provided qualitative insights into the effort required to effectively integrate the tools into students' workflows.

To assess the overall usefulness of these tools, participants provided a holistic rating of their experience throughout the project. The survey also explored participants' behavioral intentions, including their openness to using AI tools in future projects.

\begin{enumerate}
    \setcounter{enumi}{2}

    % Final Section
    \item \textit{Behavioral intentions and final reflections:}
        \begin{enumerate}
            \item \textit{``How do you rate the overall usefulness of AI tools in the programming pro\-ject?''}
            
            Scale: 1 (not helpful) to 5 (very helpful).
            \item \textit{``Would you use AI tools in future programming pro\-jects?''}
            
            Scale: 1 (very unlikely) to 5 (very likely).
            \item \textit{``Could you have achieved similar results without AI?''}
            
            Options: Yes (without extra effort), Yes (with more effort), No (AI was essential).
            \item \textit{``What new or additional AI tools or features would you find helpful?''}
            
            Open-ended response for suggestions and improvements.
            \item \textit{``Do you have any additional comments or feedback regarding the use of AI in your project?''}
            
            Open-ended response for qualitative feedback.
        \end{enumerate}

\end{enumerate}

%%%%%%%%%%%%%%%%%%%%%%%%%%%%%%%%%%%%%%%%%%%%%%%%%%%%%
\section{User Study Results}\label{s:userstudyresults}
%%%%%%%%%%%%%%%%%%%%%%%%%%%%%%%%%%%%%%%%%%%%%%%%%%%%%
Of the 38 participants in this study, 34 used AI tools during the programming project, resulting in an adoption rate of 89.4\%. The remaining four participants all reported that they did not use AI tools, citing either a lack of need or personal reservations. In addition, one participant stated that they were unaware of the AI tools available. 

\begin{figure}[ht]
  \centering
  \includegraphics[width=\textwidth]{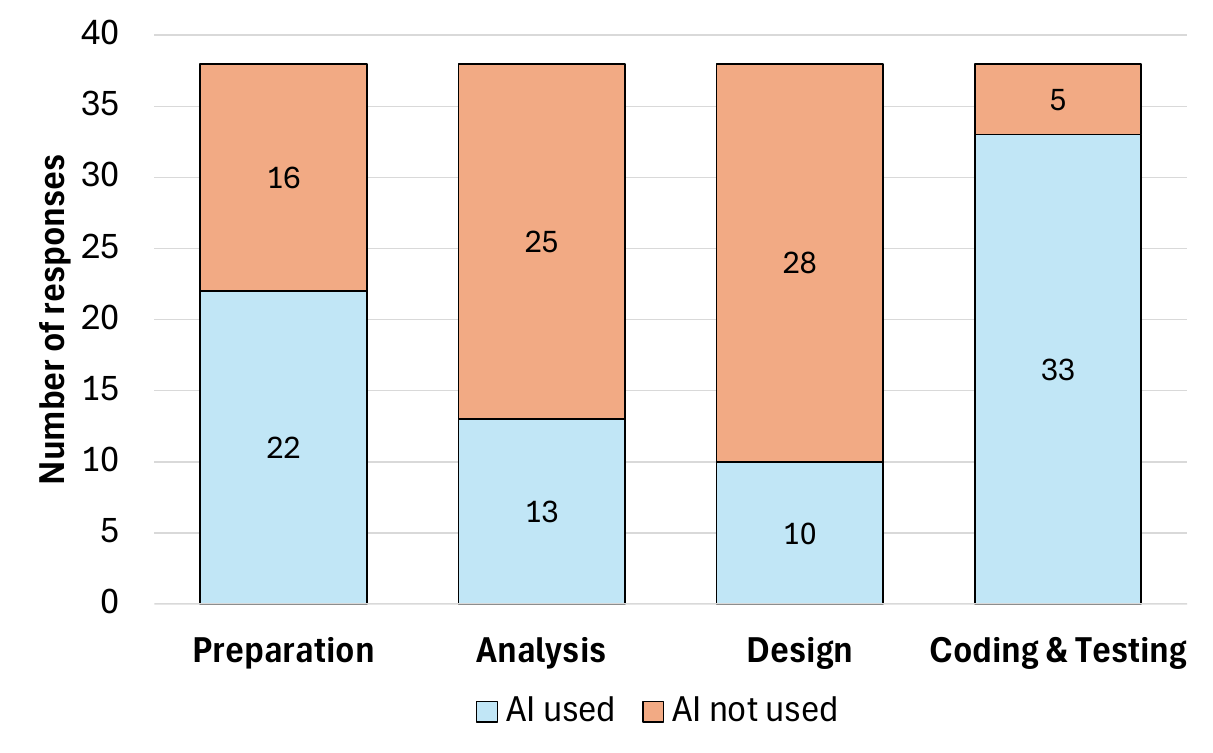}
  \caption{AI tool adoption across the project phases.}
  \label{f:usage}
\end{figure}

As shown in Fig.~\ref{f:usage}, the use of AI tools varied significantly across the four phases of the programming project. During the \preparation phase, 22 participants used AI tools, while 16 did not. In the \analysis phase, the use of AI dropped slightly, with 13 participants using the tools and 25 not using them. The \design phase saw the lowest adoption rate, with only 10 participants using AI tools, compared to 28 who abstained. Conversely, the \coding phase had the highest adoption rate, with 33 participants using AI tools and only 5 opting out. Interestingly, \chatgpt was used by all students who used AI tools, regardless of the phase. The next most commonly used tool was GitHub's \textsc{Copilot}, especially in the coding phase. Other tools were rarely used. 

Below we take a closer look at the role of AI in each phase.

\begin{figure}[ht]
  \centering
  \includegraphics[width=\textwidth]{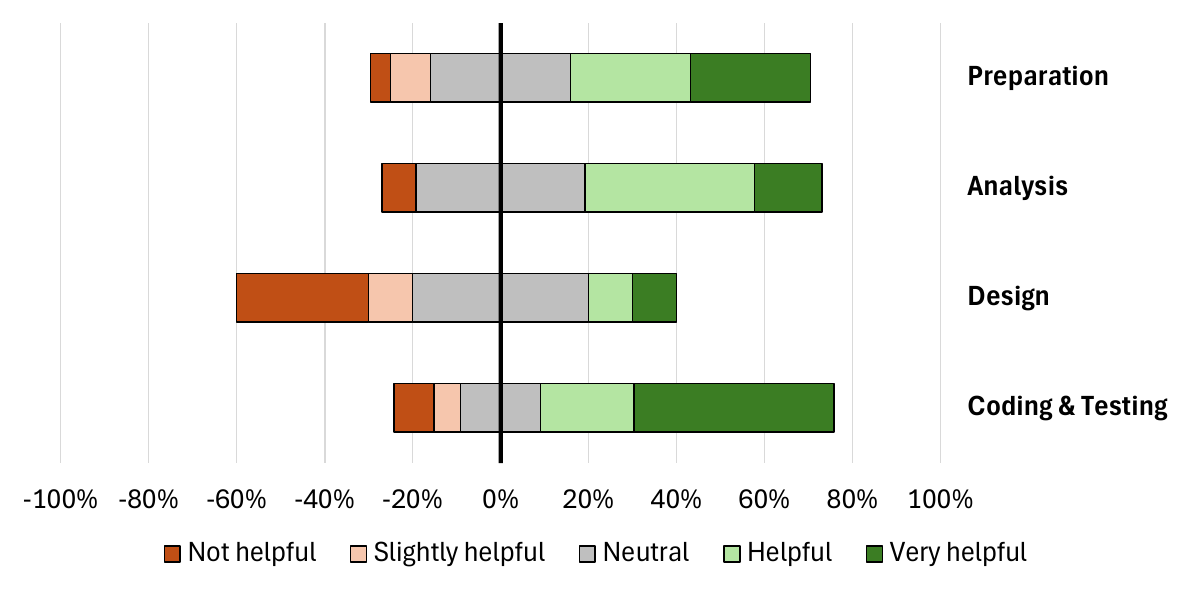}
  \caption{AI tool helpfulness across the project phases.}
  \label{f:helpfulness}
\end{figure}

Figure~\ref{f:helpfulness} illustrates the perceived helpfulness of AI tools across the four project phases using a Likert scale visualization. During both the \preparation and \coding phases, a significant proportion of participants rated AI tools as ``helpful'' or ``very helpful''.

In contrast, the \design phase saw not only the lowest adoption rates, but also the most critical ratings, with ``neutral'' and ``not helpful'' responses dominating. This is likely due to the limitations of most AI tools in generating syntactically correct diagrams, such as class or sequence diagrams, which are heavily used in this phase.

Overall, the students showed a willingness to incorporate AI into their workflows and actively attempted to complete the project tasks using AI tools. However, the nature of the tasks and the results varied depending on the phase of the project.

%%%%%%%%%%%%%%%%%%%%%%%%%%%%%%%%%%%%%%%%%%%%%%%%%%%%%
\subsection{Preparation Phase}
%%%%%%%%%%%%%%%%%%%%%%%%%%%%%%%%%%%%%%%%%%%%%%%%%%%%%
During the \preparation phase, 19 responses were collected about the tasks that participants completed using AI tools. Of these, 18 responses were categorized into specific task groups, while one response was too vague to be included in the analysis. The categorized responses highlighted different ways in which students used AI tools in their work, demonstrating both the diversity of their applications and the challenges they faced.

The most commonly reported task (n=9) was comprehension. Students relied on AI tools to understand an existing code base that formed the basis of their later project. This included working with the \jme, design patterns, and architectural choices, which required the use of AI to explain code snippets and clarify technical concepts.
However, comprehension tasks were also frequently associated with challenges. Of the nine participants who used AI for comprehension, six reported encountering ``inaccurate results'' and two mentioned ``missing context''. These problems occurred when the AI did not relate to the existing code base or was not trained on domain-specific frameworks such as \jme.

The code generation tasks (n=4) were closely related, as students were tasked with implementing additional features into the existing project. Of these participants, three noted issues with ``inaccurate results'' that required significant corrections to the AI-suggested code snippets. This issue highlights a broader limitation of AI tools in producing reliable code when the underlying logic or requirements are highly specific.

Similarly, code improvement (n=3) emerged as a recurring theme, where students used AI to incorporate feedback from their supervisors into their solutions. However, two participants reported frustration with generic or irrelevant suggestions that could not be integrated into the existing code.

The use of AI for documentation (n=3) was also notable, consistent with staff recommendations to use AI to create and refine code documentation. Although documentation tasks were relatively straightforward, one participant reported challenges with ``inaccurate results''. Debugging (n=2) was surprisingly underrepresented, despite being an introductory task for this phase, where students were asked to identify bugs in the provided code base. Both participants who used AI for debugging noted difficulties stemming from the AI's inability to understand the larger project scope, resulting in ineffective or misleading suggestions. Finally, one participant mentioned image generation (n=1), although it remains unclear what specific artifact was produced.

These findings highlight the multiple roles that AI tools played during the \preparation phase, but also reveal significant limitations in their contextual understanding and output reliability when working with unfamiliar code. The recurring issues of inaccuracy, lack of context, and overly generic suggestions show that while AI tools can improve productivity, they require careful oversight and integration into specific workflows.

%%%%%%%%%%%%%%%%%%%%%%%%%%%%%%%%%%%%%%%%%%%%%%%%%%%%%
\subsection{Analysis Phase}
%%%%%%%%%%%%%%%%%%%%%%%%%%%%%%%%%%%%%%%%%%%%%%%%%%%%%
During the \analysis phase, 13 participants reported using AI tools, with 11 providing specific tasks that were categorized into five different groups. None of the responses had to be discarded due to vague or uninterpretable responses.

Requirements specification (n=8) was the most frequently reported task. In this task, students created their own version of the requirements document, outlining how the game to be developed should work. We believe that AI helped with these tasks because text generation is one of its strengths, and many AI systems are familiar with popular games that students adapted into software during the project. Since no complaints were reported, this seems to have been a successful use of AI tools to help formulate precise and comprehensive requirements.
This was followed by use case generation (n=4), where AI tools helped create structured lists that outlined the use case's ID, priority, and key details such as purpose, actors, and conditions. This repetitive and systematic task benefited from AI by streamlining the process. However, one complaint was that the results were unreliable and required manual corrections.

Documentation tasks (n=4) focused primarily on creating user manuals. These required, among other things, detailed descriptions of the game's components, controls, and gameplay elements. As this is also a text-heavy task, it benefited from the same strengths of the AI as those observed in the requirements specification mentioned above. There were no complaints.

Diagram generation (n=2) in this phase attempts to generate class diagrams that are simple in design and complexity, as they primarily outline the preliminary classes and structure of the project. However, both mentions were accompanied by complaints that the diagrams were not correct or detailed enough. The AI's difficulty in generating meaningful diagrams was quickly apparent to the students.
Finally, GUI mockups (n=2) were used to design and present the user interface at this early stage, collaborating with the ``customer'' on how the program would look. The strength of AI in generating images definitely played a role here, allowing students to visualize and iterate on their designs. Despite the low number of mentions, several groups used AI-generated images in their projects, which was evident in the final presentations.

While only about one-third of participants reported using AI in this phase, the overall helpfulness was rated positively and few problems were reported. Text-heavy tasks such as requirements specification and documentation clearly benefited from the strengths of AI tools.
Challenges such as lack of detail and unreliable output underscore the importance of critically evaluating and refining AI-generated output to align with project requirements.

%%%%%%%%%%%%%%%%%%%%%%%%%%%%%%%%%%%%%%%%%%%%%%%%%%%%%
\subsection{Design Phase}
%%%%%%%%%%%%%%%%%%%%%%%%%%%%%%%%%%%%%%%%%%%%%%%%%%%%%
Of the 10 participants who reported using AI tools, nine provided specific tasks that were categorized into four groups. Again, no response had to be discarded due to an uninterpretable response.

The most commonly reported task was creating diagrams and charts (n=5), which in this phase included flowcharts, class, package, sequence, and state diagrams. These artifacts are meant to visualize the structural and behavioral aspect of the software, and serve as a transition from conceptual work to concrete implementation. However, and as reported in the previous phase, this may be related to the fact that two participants reported insufficient diagrams and one student specifically mentioned the lack of content in the generated diagram.

Evaluating and comparing design approaches (n=4) was another key task category, which involved evaluating different design strategies to select effective solutions, such as thick client and thin client models in a network-based architecture. One participant reported the challenge of designing a prompt with the right amount of context to elicit a sufficient response. 

Optimizing design prototypes (n=3) was also mentioned, where students asked AI tools to improve their ideas. Participants did not report any problems with this task.

Finally, creating images (n=1) was mentioned because one student reportedly continued to improve GUI elements.
No problems were reported with this task.

While most of the reported tasks were completed without reported problems, the results must be considered in light of the relatively low adoption of AI tools. Only 10 of the 38 students who participated in the study reported using them.
Furthermore, as shown in Fig.~\ref{f:helpfulness}, participants rated the helpfulness of AI at this stage the lowest, which may be due to the high demand for diagram-based artifacts.

%%%%%%%%%%%%%%%%%%%%%%%%%%%%%%%%%%%%%%%%%%%%%%%%%%%%%
\subsection{Coding \& Testing Phase}
%%%%%%%%%%%%%%%%%%%%%%%%%%%%%%%%%%%%%%%%%%%%%%%%%%%%%
This phase saw the highest adoption of AI tools, with 33 of 38 participants reporting their use. All 33 participants identified at least one task where AI provided assistance. These were grouped into 6 tasks. However, problems were prevalent, with 27 participants reporting difficulties and 24 citing one or more challenges in their feedback.

The tasks of Code Generation, Code Improvement, and Code Understanding, which were grouped together due to common themes and frequent concurrent mentions (n=28), were widely used by participants. Because this phase was primarily focused on implementing the designs from the previous phase, students reported a variety of challenges. The most commonly reported challenge, cited by 10 participants, was incorrect code. The generated code often did not integrate with the existing code base, either because of logical errors or because existing classes and methods were overlooked or misinterpreted. 

One participant specifically mentioned encountering ``hallucinations'' where the AI generated incorrect code that relied on framework methods that did not exist. Six participants noted difficulties because some AI models were not trained on domain-specific frameworks or libraries, such as \jme or \textsc{Lemur}, resulting in output that was incompatible with their projects. Three participants reported that AI tools struggled to handle the complexity of their projects because they could not provide the entire code base or large numbers of classes to the AI tool. 

This resulted in output that conflicted with existing code and did not integrate seamlessly. Finally, two participants highlighted problems with prompting, attributing their struggles to an inability to effectively communicate their requirements to the AI. Similar challenges were observed across tasks, particularly in code understanding and improvement efforts, as the AI performed better with standard Java but often failed with specialized frameworks.

Documentation tasks (n=24) were also among the most frequently reported, with participants using AI tools to create or refine their project documentation. At this stage, documentation primarily involved the creation of student-written \textsc{Javadocs} for the classes. Encouraged by the staff, many students found the AI to be particularly effective in analyzing the methods and generating comprehensive and contextual documentation. Only one participant reported encountering an erroneous AI-generated document, indicating that the tools performed well in this context.

Debugging tasks (n=20) were closely related to code generation, as both required AI tools to work within existing code bases and project frameworks. Five participants reported that the AI struggled with unfamiliar frameworks or libraries, such as \jme, echoing similar issues observed with code generation. Three participants reported incorrect output during debugging, including one instance where the AI ``hallucinated'' by generating nonsensical code that was incompatible with the context.
Other cases involved solutions that did not address the actual problem or conflicted with the existing implementation, complicating integration efforts. One participant also noted difficulties with prompting, particularly in framing complex code problems or generating contextually relevant methods.

Test generation (n=7) was another area where AI tools were used, specifically to test the model created within the \textit{Model-View-Controller} architecture. While generally effective, some problems were reported. One participant noted that poorly defined tasks for the AI resulted in outputs that did not meet the intended goals. Another participant observed that in some cases the AI produced nonsensical tests.
Finally, general questions (n=1) and music generation (n=1) were reported, but no problems with these tasks were documented.

While the \coding phase saw the highest adoption of AI tools among participants, it also revealed significant challenges, particularly with tasks such as code generation, debugging, and improving existing implementations. The tools proved valuable for automating repetitive tasks and generating initial designs, but recurring problems such as incorrect output, difficulties with domain-specific frameworks such as \jme, and limitations in handling complex code bases underscored the need for careful oversight. Despite these challenges, tasks such as documentation and test generation demonstrated the potential of AI tools to improve productivity when clear prompts and well-defined workflows were used. These findings underscore that while AI tools can significantly assist the \coding phase, their effective use requires both user expertise and improvements in tool design, particularly for specialized contexts.

%%%%%%%%%%%%%%%%%%%%%%%%%%%%%%%%%%%%%%%%%%%%%%%%%%%%%
\subsection{Final Observations}
\begin{figure}[ht]
  \centering
  \includegraphics[width=\textwidth]{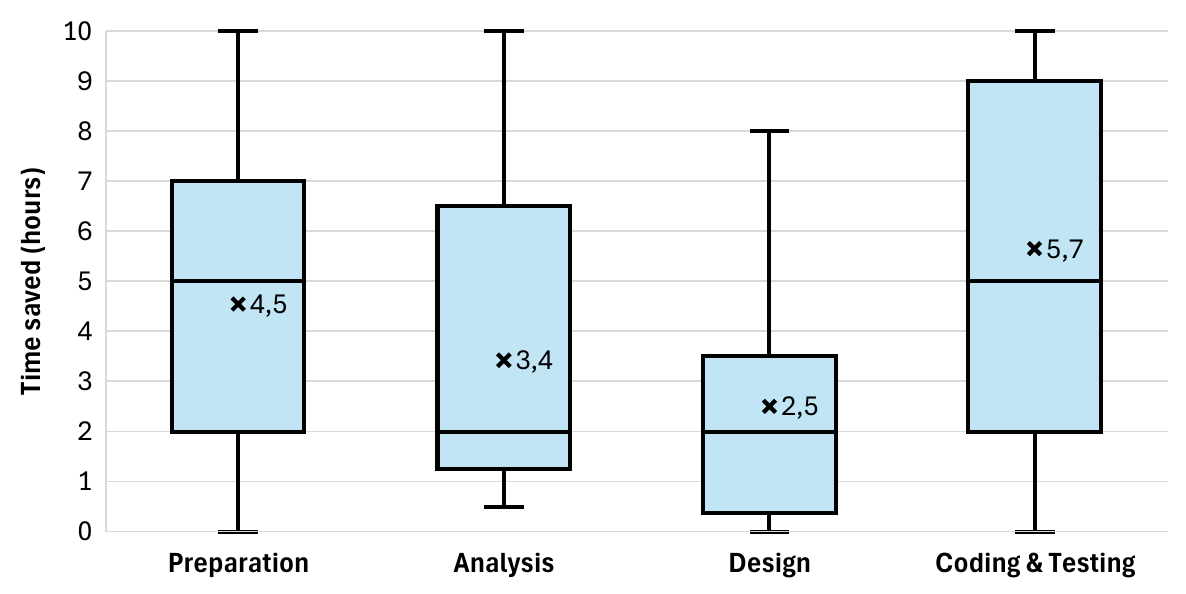}
  \caption{Self-reported time savings by project phase.}
  \label{f:time}
\end{figure}

%%%%%%%%%%%%%%%%%%%%%%%%%%%%%%%%%%%%%%%%%%%%%%%%%%%%%
In addition to analyzing specific project phases, participants were asked if they believed they could have achieved the same results without the use of AI tools. Seven participants indicated that they could have achieved similar results without AI, while 24 indicated that they could have done so, but with significantly more effort. Notably, three participants reported that AI was critical to their success. These responses underscore the significant role AI played in assisting participants during the project. While some were confident that they could achieve similar results without AI, the majority acknowledged the significant time and effort saved by integrating these tools into their workflows, as shown in Fig.~\ref{f:time}. A smaller but notable group felt that AI was essential.

\begin{figure}[ht]
  \centering
  \includegraphics[width=\textwidth]{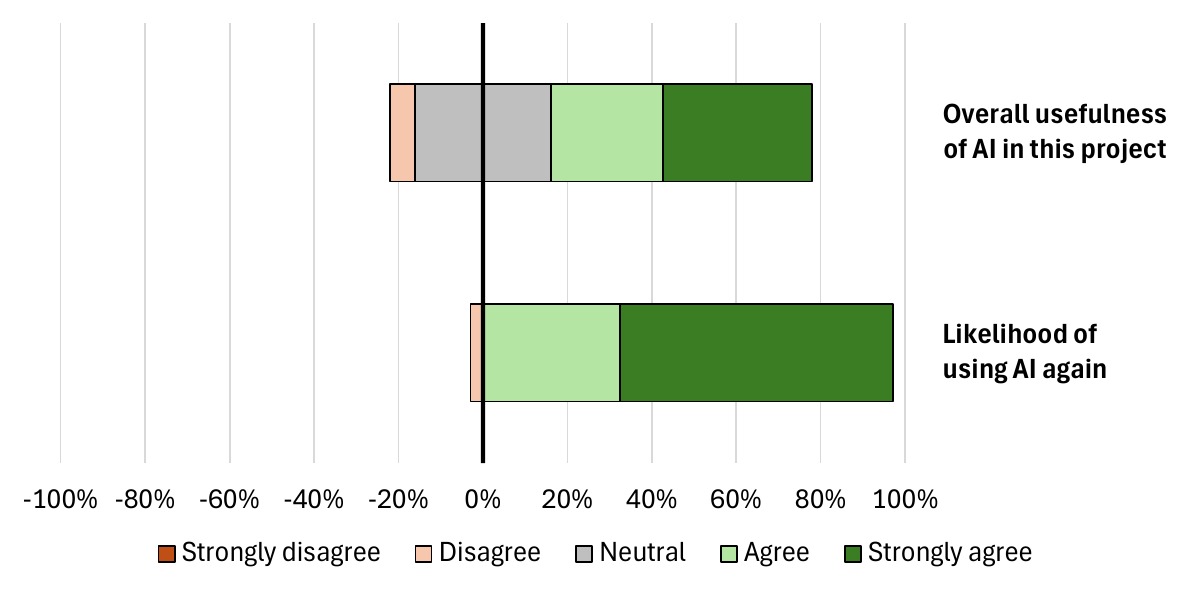}
  \caption{Perceived usefulness of AI in the project and likelihood of future use.}
  \label{f:usefulness}
\end{figure}

Interestingly, as shown in Fig.~\ref{f:usefulness}, the overwhelmingly positive perception of the role of AI contrasts with the problems reported throughout the study. Self-reported time savings varied by project phase, with the highest savings in the \coding phase and the lowest in the \design phase. However, participants frequently encountered challenges such as inaccurate or incomplete output, difficulties integrating AI-generated code with domain-specific frameworks such as \jme, and the inability of AI tools to handle complex project contexts. 

Despite these difficulties, respondents were generally positive about AI. The majority recognized the potential for AI to improve productivity, particularly for repetitive or structured tasks such as documentation and test generation. The high likelihood that respondents will use AI tools again in future projects underscores their willingness to integrate AI into their workflows despite its limitations.

%%%%%%%%%%%%%%%%%%%%%%%%%%%%%%%%%%%%%%%%%%%%%%%%%%%%%
\section{Case Study: Repository-aware LLM Assistant}
\label{s:ext-design}
%%%%%%%%%%%%%%%%%%%%%%%%%%%%%%%%%%%%%%%%%%%%%%%%%%%%%
The preceding user study showed that while students found generative AI tools helpful, especially for documentation, requirements, and other text-based tasks, they struggled when applying them to code comprehension and implementation. AI-generated code often lacked project awareness, leading to inaccuracies and integration issues. These findings suggest that effective AI support in software projects depends on contextual grounding, i.e., the ability to reference project artifacts directly.

To address this, we develop a repository-aware LLM assistant that retrieves relevant documentation and source code from the project repository before generating responses. This grounding aims to improve the accuracy, coherence, and practical usefulness of AI outputs. The system runs locally to ensure data control, reproducibility, and comprehensive traceability of interactions.
The design continues our broader goal of advancing context-sensitive AI support for software engineering education, here realized as a solution-oriented assistant rather than a tutor.

This section outlines the architecture and local deployment of the assistant, which together form the technical foundation for the evaluation presented in the following sections.

%%%%%%%%%%%%%%%%%%%%%%%%%%%%%%%%%%%%%%%%%%%%%%%%%%%%%
\subsection{Architecture}
%%%%%%%%%%%%%%%%%%%%%%%%%%%%%%%%%%%%%%%%%%%%%%%%%%%%%
Our system is designed for the programming project environment, focusing on the \preparation phase where students work with and extend the existing \battleship code base. Typical tasks in this phase include adding sound effects, introducing new game items, or extending model components with corresponding behavior and visualization. The challenges students reported when using AI—such as limited code understanding and weak integration with existing artifacts—motivate an architecture that grounds model responses in project code rather than relying on prompt-only interaction.

We structure the assistant around a chat interface, a model-serving layer, and two complementary grounding paths. 
The first, \textit{document-level RAG}, uses project documentation and auxiliary materials as preferred knowledge sources. 
All documents are embedded for semantic retrieval so that, at query time, relevant passages can be supplied as additional context. 
This mechanism directly addresses the text-heavy tasks and missing-context issues that we observed.

The second, \textit{code-level lookup}, provides targeted access to the project repository. 
Guided by the user query and, when applicable, the document-level results, the system locates relevant classes or files in the repository tree, retrieves their content through the repository API, and can optionally extract method-level snippets for more focused reasoning. 
Repository access is strictly read-only and, for now, pinned to a fixed branch. 
This path is invoked only when a prompt explicitly references code artifacts, thereby improving integration fidelity and consistency in code-related responses.

Both grounding paths feed into a unified \textit{prompt → retrieval → generation} pipeline. This pipeline constructs OpenAI-compatible \textsc{Chat Completions} API requests that combine the user prompt with the retrieved context and submits them to the model-serving endpoint, thereby keeping the architecture model-agnostic. This enables testing of different model families or decoding parameters without modifying system interfaces. 
For auditability and qualitative analysis, the system logs all prompts, retrieved passages, repository-tool inputs and outputs, and generated responses, while routing all traffic through an HTTP proxy that captures complete request traces for inspection and debugging.

%%%%%%%%%%%%%%%%%%%%%%%%%%%%%%%%%%%%%%%%%%%%%%%%%%%%%
\subsection{Deployment}
\begin{figure}[ht]
  \centering
  \includegraphics[width=\textwidth]{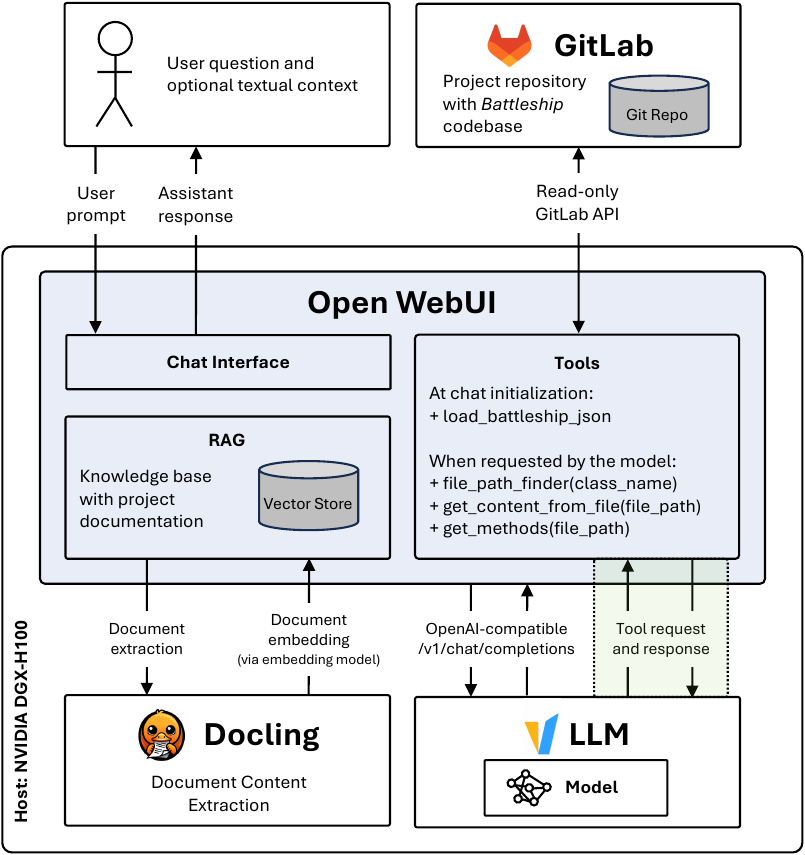}
  \caption{Local deployment stack showing the Open WebUI (chat, knowledge, tools), document grounding via Docling, model serving via vLLM (OpenAI-compatible API), and repository adapters for GitLab, all running on a DGX-H100 host.}
  \label{f:stack}
\end{figure}

%%%%%%%%%%%%%%%%%%%%%%%%%%%%%%%%%%%%%%%%%%%%%%%%%%%%%
This subsection implements the architecture and outlines the on-premises setup together with its operational controls for auditability and reproducibility. 
Figure~\ref{f:stack} provides an overview of the local system stack.

All components run as Docker containers on a dedicated on-campus NVIDIA DGX-H100 system. 
Containerization provides isolated and reproducible execution while simplifying configuration management across study runs. 
Using controlled hardware supports long-context workloads and ensures stable performance, which is critical for processing complex prompts that combine retrieved documents and multi-file code inputs.

The inference engine \textsc{vLLM}\footnote{\url{https://docs.vllm.ai}} exposes an OpenAI-compatible \textsc{Chat Completions}\footnote{\url{https://platform.openai.com/docs/api-reference/chat}} endpoint. 
We configure an extended context window to fully utilize the available GPU memory, enabling the processing of retrieved context and large source files. 
Tensor parallelism is applied across GPUs to maintain low latency for long-context workloads. 
Sampling parameters can be adjusted either through the user interface or directly within the inference engine.

The Open WebUI\footnote{\url{https://openwebui.com/}} (UI) connects to the endpoint and organizes study runs via model-chat configurations. As shown in Fig.~\ref{f:stack}, project documents are processed by a content-extraction service (Docling\footnote{\url{https://github.com/docling-project/docling}}) and embedded for semantic retrieval. The resulting knowledge base is linked to each model-chat configuration that appears as a selectable chat entry. Each configuration defines the base model, the sampling preset and model variant, the attached knowledge base with our project documentation, and the enabled tools. The overlay also includes a short system prompt instructing the assistant to answer in German and to initialize the session once by calling \texttt{load\_battleship\_json} to load the repository's class inventory. Subsequent calls of this method are disabled. At runtime, the UI assembles an OpenAI-compatible \textsc{Chat Completions} request, combining the user prompt, retrieved context, conversation history, and the enabled tools. Each request includes the tool names, natural-language descriptions, and parameter definitions. The UI submits the request to the endpoint, executes the tool calls returned by the model, and appends the tool outputs to the conversation state.

Access to the repository is provided through three dedicated tools that interface with the GitLab API. 
All calls target a fixed branch reference and rely exclusively on \texttt{GET} endpoints. 
The first tool, \texttt{file\_path\_finder}, searches the repository tree to locate relevant files. Given a class name as parameter, it returns the path of that class in the repository.
Once identified, \texttt{get\_content\_from\_file} retrieves and decodes the file content, performing minimal header cleanup when necessary. 
Finally, \texttt{get\_methods} uses Tree-Sitter–based analysis to enumerate and inspect method names within a class, enabling fine-grained reasoning over source code structure.

\begin{figure}[ht]
  \centering
  \includegraphics[width=\textwidth]{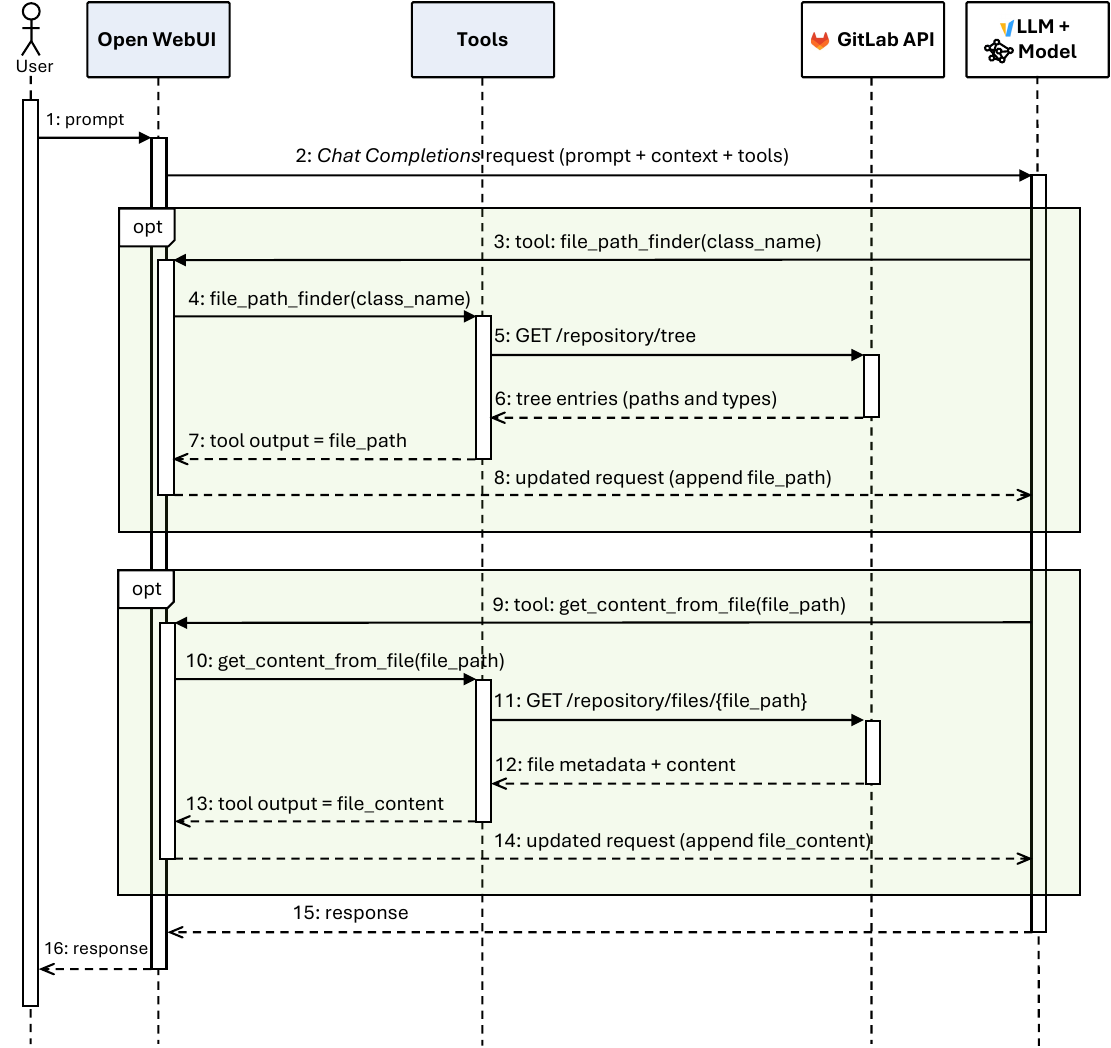}
  \caption{Exemplary tool-call pipeline: user prompt → repository path discovery (\texttt{file\_path\_finder}) $\rightarrow$ file retrieval (\texttt{get\_content\_from\_file}) $\rightarrow$ grounded response.}
  \label{f:tool-sequence}
\end{figure}

When a prompt references a specific class or method, the assistant first resolves the class path within the repository tree using \texttt{file\_path\_finder}, then retrieves the corresponding file content through \texttt{get\_content\_from\_file}, as illustrated in Fig.~\ref{f:tool-sequence}. 

The retrieved source code, together with any relevant document passages, is appended to the prompt as contextual input for the model to generate a grounded response. 
If the prompt refers to a particular method, the system first enumerates method names in the class via \texttt{get\_methods} to confirm its existence before fetching the code. 
This lookup process may be repeated across turns when additional files are referenced.

All prompts, tool inputs and outputs, and model responses are logged within the UI. 
In addition, all traffic is routed through an HTTP proxy to capture detailed traces for debugging and auditing. 
For reproducibility, the container images and the repository reference are version-pinned.

%%%%%%%%%%%%%%%%%%%%%%%%%%%%%%%%%%%%%%%%%%%%%%%%%%%%%
\section{Empirical Evaluation of the Case Study}
\label{s:ext-Study}
%%%%%%%%%%%%%%%%%%%%%%%%%%%%%%%%%%%%%%%%%%%%%%%%%%%%%
This section evaluates the effectiveness of the locally deployed, repository-aware assistant in generating project-specific responses and examines the impact of different configuration choices on response quality. 
Two main factors are considered: decoding settings, represented by sampling parameters, and variations across model families and sizes. 
The evaluation is conducted on the system stack (see Fig.~\ref{f:stack}) and focuses on two representative tasks from the \preparation phase of the \battleship programming project. 
The following describes the evaluation design, the experimental procedure, and the criteria used to assess the generated outputs.

%%%%%%%%%%%%%%%%%%%%%%%%%%%%%%%%%%%%%%%%%%%%%%%%%%%%%
\subsection{Evaluation Design}
%%%%%%%%%%%%%%%%%%%%%%%%%%%%%%%%%%%%%%%%%%%%%%%%%%%%%
The evaluation focuses on two representative tasks from the \preparation phase of the \battleship project. 
As described earlier, these tasks are intended to help students understand the existing code base and familiarize themselves with the project’s framework and overall architecture. 
The evaluation runs were conducted using the original German task formulations. For this publication, we present a faithful English translation. 
\medskip

\textit{Task 1 (3D Models for Ships)}:
``Ships are currently rendered as boxes whose length corresponds to ship size. Add a 3D model representation that assigns models based on ship length. If no model exists for a given length (or if parameters change), revert to the box representation to keep the game functional.''

This task requires an understanding of the existing, size-dependent ship rendering implementation and the introduction of a length-to-model mapping that preserves functionality when assets are missing or parameters change. 
The \battleship project includes ships of various lengths; therefore, the mapping must select the appropriate model for each length and provide a reliable fallback to ensure playability under different configurations. 
The solution should integrate into the existing rendering pipeline by identifying the relevant class, reusing existing components, and adding only minimal control logic to check model availability and activate the fallback when necessary. 
A dynamically sized box representation is already available and should serve as the default. 
This task was selected for its moderate complexity and because a 3D model for ships of length four already exists in the project, allowing the evaluation to focus on repository awareness and small, safe adaptations rather than extensive rewrites. 
Given this scope, we used the task description verbatim as the initial prompt (\textit{Prompt A1}).
\medskip

\textit{Task 2 (Background Music)}:
``The game already provides sound effects. Add suitable background music that can be enabled and disabled in the menu independently of the sound effects. In addition, the volume of the background music should be adjustable in the menu using a slider.''

This task focuses less on isolated code snippets and more on proper architectural integration. 
Background playback should be managed by a persistent application state component that participates in the game’s update lifecycle rather than being triggered ad hoc. 
It should register a dedicated music channel that is decoupled from the existing sound effects path so that user controls can operate independently. 
Within the user interface, the menu acts as the container for graphical elements: a checkbox should toggle the background music state, while a slider should bind to a small state model (e.g., volume level) maintained by the audio controller to ensure separation of view and logic. 
A clean solution therefore spans multiple classes and requires repository awareness to identify and extend the relevant code paths.

During preliminary testing on our deployed stack, early pilot runs consistently failed to produce a functional slider implementation. 
We therefore excluded the slider subtask to avoid introducing an additional UI pathway that would not affect the comparative conclusions. 
Accordingly, the evaluation was operationalized using two prompts issued within the same chat: the first established continuous background playback, and the second implemented an independent on/off control in the menu.

\textit{Prompt B1:}
``How can I implement background music in the game?''

\textit{Prompt B2:}
``How can I implement a way to enable or disable the background music independently of the sound effects? Use the class \texttt{Menu.java}.'' 

%%%%%%%%%%%%%%%%%%%%%%%%%%%%%%%%%%%%%%%%%%%%%%%%%%%%%
\subsection{Experimental Procedure}
%%%%%%%%%%%%%%%%%%%%%%%%%%%%%%%%%%%%%%%%%%%%%%%%%%%%%
Each evaluation run is defined as a new chat instance for a specific combination of model, sampling configuration, and task.
No re-runs are performed. 
To ensure comparability across runs, we fix the repository snapshot and document index, keep retrieval settings constant, and enable the same tool set in all sessions.

During each interaction, knowledge retrieval is enabled for the associated workspace knowledge base. 
Repository lookups are performed either when explicitly requested by the model or when code artifacts are required. 
The available tools are accessible to the model at every prompt and, when invoked, are executed through the user interface. 
The resulting outputs are then incorporated into the conversation state to inform subsequent model responses. 
This orchestration follows our pipeline \textit{prompt → retrieval → generation}.

Configuration effects are examined in two stages. 
In the first stage, a sampling sweep (summarized in Table~\ref{tab:sampling}) is performed using the \textit{Mistral-Large-Instruct-2411} model, while the parameters \textit{temp}, \textit{top\_p}, and \textit{min\_p} are systematically varied, thereby trading diversity against stability in code generation.

\begin{table}[ht]
\centering
\setlength{\tabcolsep}{6pt}
\renewcommand{\arraystretch}{1.30}
\caption{Sampling configurations used in the evaluation.}
\label{tab:sampling}
\begin{tabular}{lccc}
\hline
\rowcolor{lightblue}
 & temp & top\_p & min\_p \\
\hline
\multicolumn{4}{l}{\textit{Standard (no change)}}\\
Default & 1.0 & 1.0 & 0.0 \\
\hline
\multicolumn{4}{l}{\textit{Isolated change}}\\
temp 0.5 & 0.5 & 1.0 & 0.0 \\
temp 0.3 & 0.3 & 1.0 & 0.0 \\
top\_p 0.8       & 1.0 & 0.8 & 0.0 \\
top\_p 0.5       & 1.0 & 0.5 & 0.0 \\
min\_p 0.1       & 1.0 & 1.0 & 0.1 \\
min\_p 0.3       & 1.0 & 1.0 & 0.3 \\
\hline
\multicolumn{4}{l}{\textit{Combined change}}\\
temp 0.5 + min\_p 0.1 & 0.5 & 1.0 & 0.1 \\
temp 0.5 + min\_p 0.3 & 0.5 & 1.0 & 0.3 \\
temp 0.3 + min\_p 0.1 & 0.3 & 1.0 & 0.1 \\
temp 0.3 + min\_p 0.3 & 0.3 & 1.0 & 0.3 \\
top\_p 0.8 + min\_p 0.1 & 1.0 & 0.8 & 0.1 \\
top\_p 0.8 + min\_p 0.3 & 1.0 & 0.8 & 0.3 \\
top\_p 0.5 + min\_p 0.1 & 1.0 & 0.5 & 0.1 \\
top\_p 0.5 + min\_p 0.3 & 1.0 & 0.5 & 0.3 \\

\hline
\end{tabular}
\end{table}

In the second stage, a model sweep is conducted with the sampling parameters fixed at \(\textit{temp}=0.5\), \(\textit{top\_p}=0.95\), and \(\textit{min\_p}=0\), while varying the model families and sizes. These values were chosen \emph{a priori}, guided by a brief pre-study, to balance response consistency with controlled variability in generating outputs. The lineup spans multiple families and sizes, including code-specialized and general-purpose variants. To ensure comparability, the task prompts remain identical across both stages.

We evaluated the following models.

\begin{itemize}
  \item \textit{DeepSeek-Coder-V2-Instruct}\footnote{\url{https://huggingface.co/deepseek-ai/DeepSeek-Coder-V2-Instruct}} (236B parameters)
  \item \textit{Devstral-Small-2505}\footnote{\url{https://huggingface.co/mistralai/Devstral-Small-2505}} (24B parameters)
  \item \textit{Mistral-Large-Instruct-2411}\footnote{\url{https://huggingface.co/mistralai/Mistral-Large-Instruct-2411}} (123B parameters)
  \item \textit{Mistral-7B-Instruct-v0.3}\footnote{\url{https://huggingface.co/mistralai/Mistral-7B-Instruct-v0.3}} (7B parameters)
  \item \textit{Qwen3-235B-A22B}\footnote{\url{https://huggingface.co/Qwen/Qwen3-235B-A22B}} (235B parameters)
  \item \textit{Qwen3-30B-A3B}\footnote{\url{https://huggingface.co/Qwen/Qwen3-30B-A3B}} (30B parameters)
\end{itemize}

To ensure comparability, we disabled model-native \emph{function calling} (where available) and routed all tool use through the same UI-controlled path for all models.
Each configuration–task combination is executed exactly once. 
Task~1 is issued as a single prompt, whereas Task~2 uses its two-step prompt sequence within the same chat. 
Beyond these predefined steps, no restarts or re-runs are performed. 

For auditability, we record all user prompts, retrieved document passages, repository tool inputs and outputs, and model responses from the user interface logs.
These data are complemented by HTTP-level traces captured through the proxy to ensure complete transparency and reproducibility.

%%%%%%%%%%%%%%%%%%%%%%%%%%%%%%%%%%%%%%%%%%%%%%%%%%%%%
\subsection{Defect Catalog}
%%%%%%%%%%%%%%%%%%%%%%%%%%%%%%%%%%%%%%%%%%%%%%%%%%%%%
The generated responses were analyzed according to a catalog of recurrent defect types identified during the evaluation. 
The catalog includes the following categories.

\begin{enumerate}
  \item \textit{Hallucination.} 
  The model refers to or calls code artifacts that do not exist in the repository but are assumed to be available. 
  Typical cases include:
  \begin{enumerate}
    \item Referencing undefined or missing types.
    \item Invoking methods that are not defined for an existing class or type.
  \end{enumerate}

  \item \textit{Tool misuse.} 
  Incorrect invocation or handling of repository tools. 
  Examples include:
  \begin{enumerate}
    \item Using a repository tool with an incorrect target (e.g., the wrong class or method).
    \item Retrieving the correct artifact but ignoring the result in the generated response.
  \end{enumerate}

  \item \textit{Task misunderstanding.} 
  Partial or incorrect interpretation of the user’s request. 
  For instance, the model may implement back-end logic but fail to expose the corresponding functionality in the user interface.

  \item \textit{Insufficient robustness.} 
  The proposed code lacks defensive checks for missing or unloadable resources and fails to degrade gracefully under error conditions. 
  Common examples are:
  \begin{enumerate}
    \item Constructing a resource path dynamically without validating its existence.
    \item Mentioning a fallback mechanism but omitting exception handling, leaving failures unhandled.
  \end{enumerate}

  \item \textit{Use-before-initialization.} 
  Accessing objects that are not yet constructed or fully initialized. 
  Typical occurrences include:
  \begin{enumerate}
    \item Passing a \texttt{null} reference to a method expecting an instance.
    \item Invoking a method on a field before its initializer has executed.
  \end{enumerate}

  \item \textit{Duplicate variable declaration.} 
  Declaring the same identifier multiple times within the same scope, such as redeclaring a local variable inside a method body.

  \item \textit{Missing resource entry.} 
  Adding new non-code resources without registering them properly. 
  For example, adding a new UI element with a text label but omitting the corresponding localization key or value in the resource bundle.

  \item \textit{Type-mismatched comparison.} 
  Using comparison operators on operands of incompatible data types. 
  For instance, comparing a method’s enum return value with a constant from a different enumeration.

  \item \textit{Invalid invocation.} 
  Performing illegal or context-inappropriate method calls or field accesses. 
  A common case is attempting to access instance fields from a static context.

  \item \textit{Integration omission.} 
  Failing to reuse existing code paths or helper functions when extending functionality. 
  This often leads to incomplete or inconsistent behavior, such as omitting calls to necessary utility methods.

  \item \textit{Wrapper-only method.} 
  Creating a method that simply delegates to another without adding meaningful logic. 
  A typical example is calling an overloaded method with mapped parameters and immediately returning its result.

  \item \textit{Code duplication.} 
  Repeating the same logic across multiple methods with only minor variations. 
  For example, two methods may share nearly identical bodies that differ only by a constant or parameter.
\end{enumerate}

%%%%%%%%%%%%%%%%%%%%%%%%%%%%%%%%%%%%%%%%%%%%%%%%%%%%%
\section{Results of the Case Study Evaluation}
\label{s:ext-results}
%%%%%%%%%%%%%%%%%%%%%%%%%%%%%%%%%%%%%%%%%%%%%%%%%%%%%
This section presents the outcomes of the case study after the two-stage evaluation. 
In total, 42 runs were conducted, comprising 15 sampling configurations and six model variants, each evaluated across two tasks. Across both stages, the repository-aware setup consistently produced usable, repository-aligned outputs. In most runs, the generated artifacts could be integrated into the existing code base with limited follow-up corrections. This stands in contrast to the generic assistants observed in our user study, which frequently lacked sufficient project context for seamless integration.

All generated outputs were analyzed using the defect catalog, with every observed defect classified by category and, where applicable, by case variant. 
The following subsections first discuss the results of the sampling sweep, then the model sweep, and finally address the identified threats to validity.

%%%%%%%%%%%%%%%%%%%%%%%%%%%%%%%%%%%%%%%%%%%%%%%%%%%%%
\subsection{Sampling Sweep}
%%%%%%%%%%%%%%%%%%%%%%%%%%%%%%%%%%%%%%%%%%%%%%%%%%%%%
Table~\ref{tab:sampling-result} summarizes the defects observed across the 15 sampling configurations. 

\begin{table}[tb]
\caption{Sampling sweep results. The (upper) blue entries show defects from Task~1, and the (lower) red entries show defects from Task~2. 
Letters \textit{a} and \textit{b} correspond to the case variants defined in the defect catalog. 
Entries marked with \texttt{x} indicate a defect category without a specific case, while \texttt{x2} and \texttt{x3} denote repeated occurrences. 
Empty cells indicate that no defects of the corresponding category were observed.}
\label{tab:sampling-result}
\centering
\setlength{\tabcolsep}{6pt}
\renewcommand{\arraystretch}{1.60}
\begin{tabular}{|c|c|c|*{12}{c|}}
\hline
\rowcolor{lightblue}
\rotatebox{90}{\textbf{temp}} & 
\rotatebox{90}{\textbf{top\_p}} & 
\rotatebox{90}{\textbf{min\_p}} &
\rotatebox{90}{Hallucination} &
\rotatebox{90}{Tool misuse} &
\rotatebox{90}{Task misunderstanding} &
\rotatebox{90}{Insufficient robustness} &
\rotatebox{90}{Use-before-initialization} &
\rotatebox{90}{Duplicate variable declaration } &
\rotatebox{90}{Missing resource entry} &
\rotatebox{90}{Type-mismatched comparison } &
\rotatebox{90}{Invalid invocation} &
\rotatebox{90}{Integration omission} &
\rotatebox{90}{Wrapper-only method} &
\rotatebox{90}{Code duplication}\\
\hline
\multicolumn{15}{|l|}{\textit{Standard (no change)}}\\
\hline
1.0 & 1.0 & 0.0 & \scell{}{b} &  &  &  & \scell{x}{} &  & \scell{}{x} &  &  & \scell{x}{} &  &  \\\hline
\multicolumn{15}{|l|}{\textit{Isolated change}}\\
\hline
\textbf{0.5} & 1.0 & 0.0 &  &  &  &  &  &  & \scell{}{x} &  &  &  &  &  \\\hline
\textbf{0.3} & 1.0 & 0.0 & \scell{}{b} & \scell{}{a} &  &  &  &  & \scell{}{x} &  &  &  &  &  \\\hline
1.0 & \textbf{0.8} & 0.0 & \scell{a}{b} &  &  & \scell{a}{} &  &  & \scell{}{x} &  &  &  &  &  \\\hline
1.0 & \textbf{0.5} & 0.0 & \scell{a}{b} &  &  &  &  &  & \scell{}{x} &  &  &  &  &  \\\hline
1.0 & 1.0 & \textbf{0.1} & \scell{}{b x2} &  &  &  & \scell{x}{} &  & \scell{}{x} &  &  &  &  & \scell{x}{} \\\hline
1.0 & 1.0 & \textbf{0.3} & \scell{}{b x2} &  &  &  &  &  & \scell{}{x} &  &  &  & \scell{x}{} &  \\\hline

\multicolumn{15}{|l|}{\textit{Combined change}}\\
\hline
\textbf{0.5} & 1.0 & \textbf{0.1} & \scell{}{b} &  &  & \scell{a}{} & \scell{}{b} &  & \scell{}{x} & \scell{}{x} &  &  &  &  \\\hline
\textbf{0.5} & 1.0 & \textbf{0.3} & \scell{}{b} &  &  &  &  &  & \scell{}{x} &  &  & \scell{x}{} &  &  \\\hline
\textbf{0.3} & 1.0 & \textbf{0.1} & \scell{}{b} & \scell{}{a} &  &  &  &  & \scell{}{x} &  &  & \scell{x}{} &  &  \\\hline
\textbf{0.3} & 1.0 & \textbf{0.3} & \scell{}{b} &  &  &  &  & \scell{x}{} & \scell{}{x} & \scell{}{x} &  &  &  &  \\\hline
1.0 & \textbf{0.8} & \textbf{0.1} & \scell{}{b x3} &  &  & \scell{a}{} &  &  & \scell{}{x} &  &  &  &  &  \\\hline
1.0 & \textbf{0.8} & \textbf{0.3} & \scell{}{b x2} &  &  &  &  &  & \scell{}{x} &  &  &  &  &  \\\hline
1.0 & \textbf{0.5} & \textbf{0.1} & \scell{}{b} &  &  &  &  &  & \scell{}{x} &  & \scell{}{x} &  &  &  \\\hline
1.0 & \textbf{0.5} & \textbf{0.3} & \scell{}{b} &  &  & \scell{a}{} &  &  & \scell{}{x} &  &  &  &  &  \\\hline

\end{tabular}
\end{table}

A clear pattern emerged with respect to \textit{Hallucination}, which was the most frequent defect, occurring in 14 out of 15 runs and clustering most prominently at \(T=1.0\). 
Even when the token selection range was narrowed using \(\textit{top\_p}\in\{0.5,\,0.8\}\) or \(\textit{min\_p}>0\), hallucinations still appeared—typically as isolated cases rather than the multiple occurrences (two or three) observed at higher temperatures.

Beyond \textit{Hallucination}, other defect categories did not appear consistently across different levels of decoding noise. 
The \textit{Missing resource entry} defect was largely unaffected by sampling variation and recurred even in otherwise clean configurations. 
Because our system logs all prompts, responses, tool operations, and retrieved document passages, we can confirm that the required resource files were present. 
This pattern indicates a retrieval-to-action gap, in which relevant context is successfully retrieved but not incorporated into the generated output.
Apart from \textit{Hallucination} and \textit{Missing resource entry}, the remaining defect observations were too sparse, given the single-run design, to support parameter-specific conclusions. 
Defects such as \textit{Type-mismatched comparison} and \textit{Invalid invocation} occurred only sporadically. 
\textit{Wrapper-only method} and \textit{Code duplication} were observed exclusively at \(T=1.0,\, \textit{top\_p}=1.0\). 
We therefore interpret these as low base-rate phenomena rather than systematic effects. 
Additional repetitions of the runs would be required to assess variance and confirm the stability of these findings.

In conclusion, the configuration \(T\approx 0.5,\ \textit{top\_p}=0\) proved to be the most stable, yielding the fewest defects across all runs. 
Configurations combining \(T=1.0\) or a positive \(\textit{min\_p}\) with additional filtering (e.g., \((1.0,1.0,0.1)\), \((1.0,0.8,0.1)\)) consistently produced multiple issues. 
Accordingly, we adopt a conservative default configuration for code-centric applications.

%%%%%%%%%%%%%%%%%%%%%%%%%%%%%%%%%%%%%%%%%%%%%%%%%%%%%
\subsection{Model Sweep}
%%%%%%%%%%%%%%%%%%%%%%%%%%%%%%%%%%%%%%%%%%%%%%%%%%%%%
Table~\ref{tab:model-result} presents the defect profiles of six models evaluated under fixed decoding parameters and identical task prompts. 
A consistent pattern was observed across model families: in the initial prompt, the models often identified the correct class but hallucinated its content. 
After being instructed to retrieve the corresponding file, the generated outputs adjusted accordingly, and the hallucinations did not reappear. 
Therefore, these early but self-corrected cases were not included in the defect catalog.

\begin{table}[tb]
\caption{Model sweep results. The (upper) blue entries show defects from Task~1, and the (lower) red entries show defects from Task~2.  
Letters \textit{a} and \textit{b} correspond to the case variants defined in the defect catalog. 
Entries marked with \texttt{x} indicate a defect category without a specific case, while \texttt{x2} denotes repeated occurrences. 
Empty cells indicate that no defects of the corresponding category were observed.}
\label{tab:model-result}
\centering
\setlength{\tabcolsep}{6pt}
\renewcommand{\arraystretch}{1.60}
\begin{tabular}{|l|*{12}{c|}}
\hline
\rowcolor{lightblue}
\textbf{Model (abbrev.)} &
\rotatebox{90}{Hallucination} &
\rotatebox{90}{Tool misuse} &
\rotatebox{90}{Task misunderstanding} &
\rotatebox{90}{Insufficient robustness} &
\rotatebox{90}{Use-before-initialization} &
\rotatebox{90}{Duplicate variable declaration } &
\rotatebox{90}{Missing resource entry} &
\rotatebox{90}{Type-mismatched comparison } &
\rotatebox{90}{Invalid invocation} &
\rotatebox{90}{Integration omission} &
\rotatebox{90}{Wrapper-only method} &
\rotatebox{90}{Code duplication}\\
\hline
DeepSeek-Coder-V2%-Instruct (236B)  
& \scell{}{b} & \scell{}{b} & & & & & \scell{}{x} & & & & & \\\hline
Devstral-Small-2505% (23.6B)        
& \scell{}{b} & & & & & & \scell{}{x} & & & & & \\\hline
Mistral-Large%-Instruct-2411 (123B) 
& \scell{}{b} & & & & & & \scell{}{x} & & & & & \\\hline
Mistral-7B%-Instruct-v0.3 (7.25B)   
& & & \scell{}{x} & & & & \scell{}{x} & & & \scell{x}{} & & \scell{x}{} \\\hline
Qwen3-235B-A22B% (235B)             
& \scell{}{b} & & & & & & & \scell{}{x} & & & & \\\hline
Qwen3-30B-A3B% (30.5B)              
& \scell{}{b x2} & & & \scell{b}{} & & & \scell{}{x} & & & & & \\\hline
\end{tabular}
\end{table}

Two notable tendencies emerged despite the single-run design. 
First, \textit{Hallucination} was not confined to a specific model family or size, as every model except \textit{Mistral-7B} exhibited at least one marked instance. 
For \textit{Mistral-7B}, the \textit{Hallucination} category remained unannotated only because the run failed to produce complete solutions for both tasks. 
Second, the \textit{Missing resource entry} defect recurred across almost all models. 
As in the sampling sweep, this appears to stem not from decoding noise but from a retrieval-to-action gap. 
Again, the system logs confirmed successful retrieval of the relevant documentation, yet the required property entry was never written.

Among the large models, \textit{DeepSeek-Coder-V2-Instruct} ultimately solved both tasks with repository-aware code, although it required an explicit follow-up before identifying the correct classes. 
\textit{Mistral-Large-Instruct-2411} produced a clean design and demonstrated coherent component linking. 
\textit{Qwen3-235B} generated the most comprehensive rationale and was the only model to recognize the need to update the resource file, yet it still required corrections to achieve a fully functional solution.

Among the mid-sized models, \textit{Devstral-Small-2505} performed well relative to its size: it successfully completed the mapping task and, uniquely, identified persistent user preferences in Task~2. 
\textit{Qwen3-30B} completed the ship-model mapping but failed to connect all components. 
The smallest model, \textit{Mistral-7B}, proved unsuitable for this context: multiple follow-up interactions were necessary, partial code was produced without integration into the existing workflow, and the run did not reach completion.

Two practical observations can be drawn. 
First, model size alone did not predict success: for instance, the 23.6B \textit{Devstral-Small-2505} produced the most usable outputs despite its medium size. 
In contrast, very large models provided more detailed justifications but did not consistently achieve better integration. 
Second, the recurring cross-model defects indicate that improving tool utilization has a greater impact than switching model families when operating under similar constraints.

%%%%%%%%%%%%%%%%%%%%%%%%%%%%%%%%%%%%%%%%%%%%%%%%%%%%%
\subsection{Threats to Validity}
%%%%%%%%%%%%%%%%%%%%%%%%%%%%%%%%%%%%%%%%%%%%%%%%%%%%%
This study is limited by a single-run protocol per configuration and task, which constrains reproducibility and the stability of observed effects in a stochastic setting. 
The analysis relies on manual coding against a predefined defect catalog. 
Although logs capture prompts, tool invocations, and retrieved passages, the absence of a standardized benchmark introduces subjectivity and potential coder bias. 
External validity is constrained by the narrow scope of two tasks drawn from a single project; other architectures or task types may yield different patterns. 
Tooling choices also influence the results: we standardized the same UI-controlled repository tools across all models and disabled model-specific native tool integrations to ensure comparability, which may underrepresent the capabilities of models with stronger native tooling. 
Finally, the retrieval path employs a relatively simple keyword-based search and conservative filtering. 
Broader repository searches were excluded after early experiments due to latency and context-window limitations, suggesting that some integration failures may result from retrieval granularity rather than intrinsic model behavior. 
Time and operational constraints further limited the number of models and repetitions that could be tested.

%%%%%%%%%%%%%%%%%%%%%%%%%%%%%%%%%%%%%%%%%%%%%%%%%%%%%
\section{Conclusions and Future Work}\label{s:conclusion}
%%%%%%%%%%%%%%%%%%%%%%%%%%%%%%%%%%%%%%%%%%%%%%%%%%%%%
This paper presented an empirical study on the use of generative and repository-aware AI in student software development projects. Conducted within a university programming project, the study combined quantitative insights from student experiences with qualitative observations of a locally deployed, retrieval-augmented LLM assistant. The objective was to examine how AI tools can effectively support software engineering education and how their integration can be improved beyond general-purpose usage.

The results show that students perceive generative AI as valuable, particularly for text-heavy and structured tasks such as documentation, requirements formulation, and test generation. However, persistent challenges remain in code comprehension and implementation, where AI-generated outputs often fail to align with the project’s structure and dependencies. This integration weakness, previously identified by \cite{BorghoffMS25}, is substantially mitigated in our repository-aware setup through the use of project context and dedicated retrieval tools. Grounding model responses in actual repository artifacts enables more coherent, contextually aligned, and reproducible outputs.

The study also shows that the advantages and limitations of AI assistance vary across the different phases of the software project. Tasks that rely mainly on text benefit most from generative support, whereas diagram and design related artifacts, such as class and sequence diagrams, are still not well supported. The focus on text-based work therefore remains consistent, while the extension of AI support to visual and conceptual design activities continues to be an important area for future investigation.

Qualitative analysis of the repository-aware assistant revealed recurring model behaviors such as context drift, tool misuse, and occasional hallucinations. These findings emphasize the need for more precise retrieval mechanisms, better prompt control, and refined tool integration. Nonetheless, the inclusion of repository context already improves both the accuracy and interpretability of AI-assisted results, transforming the interaction from isolated prompt--response exchanges into a more integrated and traceable workflow.

In summary, the study demonstrates that combining generative capabilities with repository-grounded retrieval can meaningfully strengthen AI assistance in software engineering education. Future work will focus on refining this approach toward a reflective, tutoring-oriented mode that supports reasoning rather than replacement, and on extending contextual grounding to include visual and design artifacts. 

Our long-term goal for the programming project is to develop an AI system that acts as a tutor, assisting students with their programming assignments in an adaptive and pedagogically meaningful way. We envision two complementary approaches: one in which the AI first generates a provisional solution from the student's context and then provides targeted guidance, and another in which it operates without first forming a solution hypothesis, encouraging students to engage in exploratory problem-solving. Developing case studies for both scenarios will allow us to evaluate their educational effectiveness and make an informed decision on how best to integrate such an AI tutor into the curriculum. By embedding these context-aware tutoring capabilities into future iterations of the project, we aim to foster deeper conceptual understanding, reflective learning, and responsible use of AI in software engineering education.

\backmatter

%%%%%%%%%%%%%%%%%%%%%%%%%%%%%%%%%%%%%%%%%%%%%%%%%%%%%%%%%%%%%
\bmhead{Acknowledgements}
%%%%%%%%%%%%%%%%%%%%%%%%%%%%%%%%%%%%%%%%%%%%%%%%%%%%%%%%%%%%%
We would like to thank our best students for their entertaining and creative game projects, which we continue to enjoy testing at the Institute’s traditional Christmas event. 
We also wish to express our sincere gratitude to our students Vincent Bongiorno and Lucien Heller for their outstanding thesis work, whose ideas and technical contributions have influenced this paper. 
Their efforts have significantly enriched both the system design and the methodological approach presented here. 
Finally, we acknowledge the use of \textsc{DeepL Translator}, \textsc{DeepL Write},  \textsc{Grammarly}, and \textsc{ChatGPT 5} for language improvement and translation support. 
These tools were used to refine the text and to translate selected sections from the original German version. 
The paper, however, remains an accurate representation of the authors’ original research and contributions.

%%%%%%%%%%%%%%%%%%%%%%%%%%%%%%%%%%%%%%%%%%%%%%%%%%%%%%%%%%%%%
\section*{Declarations}
%%%%%%%%%%%%%%%%%%%%%%%%%%%%%%%%%%%%%%%%%%%%%%%%%%%%%%%%%%%%%

%%%%%%%%%%%%%%%%%%%%%%%%%%%%%%%%%%%%%%%%%%%%%%%%%%%%%%%%%%%%%
%\subsection*{Funding}
%%%%%%%%%%%%%%%%%%%%%%%%%%%%%%%%%%%%%%%%%%%%%%%%%%%%%%%%%%%%%
%Open access funding was enabled by Projekt DEAL and organized by the University of the Bundeswehr Munich.

%%%%%%%%%%%%%%%%%%%%%%%%%%%%%%%%%%%%%%%%%%%%%%%%%%%%%%%%%%%%%
\subsection*{Ethical Approval}
%%%%%%%%%%%%%%%%%%%%%%%%%%%%%%%%%%%%%%%%%%%%%%%%%%%%%%%%%%%%%
The study involving humans was approved by the Ethics Committee (IRB) at University of the Bundeswehr Munich under \textit{Ethics Committee Approval -- EK UniBw M 25-50}.
The study was conducted in accordance with local legislation and institutional requirements for data protection.

%%%%%%%%%%%%%%%%%%%%%%%%%%%%%%%%%%%%%%%%%%%%%%%%%%%%%%%%%%%%%
\subsection*{Informed Consent}
%%%%%%%%%%%%%%%%%%%%%%%%%%%%%%%%%%%%%%%%%%%%%%%%%%%%%%%%%%%%%
A total of 49 students were enrolled in the course, of whom 38 voluntarily participated in the user study. 
All participants signed a written consent form agreeing to take part in the study. 
They also consented to the anonymized use of their collected data for research and publication purposes. 
Written consent was obtained from all participants prior to the start of the study.

%%%%%%%%%%%%%%%%%%%%%%%%%%%%%%%%%%%%%%%%%%%%%%%%%%%%%%%%%%%%%
\subsection*{Author Contributions}
%%%%%%%%%%%%%%%%%%%%%%%%%%%%%%%%%%%%%%%%%%%%%%%%%%%%%%%%%%%%%
All authors made equal, substantial, direct, and intellectual contributions to the work and approved it for publication.

%%%%%%%%%%%%%%%%%%%%%%%%%%%%%%%%%%%%%%%%%%%%%%%%%%%%%%%%%%%%%
\subsection*{Code and Data Availability}
%%%%%%%%%%%%%%%%%%%%%%%%%%%%%%%%%%%%%%%%%%%%%%%%%%%%%%%%%%%%%
The original German task formulations and the datasets generated and/or analyzed during the current study are available from the corresponding authors.

%%%%%%%%%%%%%%%%%%%%%%%%%%%%%%%%%%%%%%%%%%%%%%%%%%%%%%%%%%%%%
\subsection*{Conflict of Interest}
%%%%%%%%%%%%%%%%%%%%%%%%%%%%%%%%%%%%%%%%%%%%%%%%%%%%%%%%%%%%%
The authors declare that there is no conflict of interest regarding the publication of this paper.

%%%%%%%%%%%%%%%%%%%%%%%%%%%%%%%%%%%%%%%%%%%%%%%%%%%%%%%%%%%%%
%%%%%%%%%%%%%%%%%%%%%%%%%%%%%%%%%%%%%%%%%%%%%%%%%%%%%%%%%%%%%
%\bibliography{bibliography}   % name your BibTeX data base

%%%%%%%%%%%%%%%%%%%%%%%%%%%%%%%%%%%%%%%%%%%%%%%%%%%%%%%%%%%%%
%%%%%%%%%%%%%%%%%%%%%%%%%%%%%%%%%%%%%%%%%%%%%%%%%%%%%%%%%%%%%

\end{document}